\documentclass[sigconf, nonacm]{acmart}
\newcommand\vldbdoi{XX.XX/XXX.XX}
\newcommand\vldbpages{XXX-XXX}
\newcommand\vldbvolume{18}
\newcommand\vldbissue{8}
\newcommand\vldbyear{2025}
\newcommand\vldbauthors{\authors}
\newcommand\vldbtitle{\shorttitle} 
\newcommand\vldbavailabilityurl{https://github.com/M2oDA-Lab/Roq}
\newcommand\vldbpagestyle{empty} 

\usepackage{algorithm}
\usepackage{algorithmic}
\usepackage{subcaption}
\usepackage{appendix}

\usepackage{booktabs}    
\usepackage{multirow}
\usepackage{balance}

\AtBeginDocument{%
  \providecommand\BibTeX{{%
    \normalfont B\kern-0.5em{\scshape i\kern-0.25em b}\kern-0.8em\TeX}}}

\newtheorem{theorem}{Theorem}[section]

\newtheorem{example}{Example}[section]
\newtheorem{definition}{Definition}[section]

\begin{document}

\title{Robust Plan Evaluation based on Approximate Probabilistic Machine Learning}


\author{Amin Kamali}
\email{skama043@uottawa.ca}
\affiliation{%
  \institution{University of Ottawa}
  \city{Ottawa}
  \country{Canada}
}

\author{Verena Kantere}
\email{vkantere@uottawa.ca}
\affiliation{%
  \institution{University of Ottawa}
  \city{Ottawa}
  \country{Canada}
}

\author{Calisto Zuzarte}
\email{Calisto@ca.ibm.com}
\affiliation{%
  \institution{IBM Canada Lab}
 \city{Markham}
 \country{Canada}
}

\author{Vincent Corvinelli}
\email{vcorvine@ca.ibm.com}
\affiliation{%
  \institution{IBM Canada Lab}
 \city{Markham}
 \country{Canada}
}


\begin{abstract}
  Query optimizers in RDBMSs 
  search for execution plans expected to be optimal for given queries. They use parameter estimates, often inaccurate, and make assumptions that may not hold in practice. Consequently, they may select plans that are
  suboptimal at runtime, if estimates and assumptions are not valid. Therefore, they do not sufficiently support robust query optimization. 
  Using ML 
  to improve 
  data systems 
  has shown promising results for query optimization. Inspired by this, 
  we propose \underline{Ro}bust \underline{Q}uery Optimizer, (Roq), a holistic framework based on a risk-aware learning approach. Roq includes a novel formalization of the notion of robustness in the context of query optimization and a principled approach for its quantification and measurement based on approximate probabilistic ML. It also includes novel strategies and algorithms for query plan evaluation and selection. Roq includes a novel learned cost model that is designed to predict the cost of query execution and the associated risks and performs query optimization accordingly. We demonstrate 
  that Roq provides significant improvements in robust query optimization compared with the state-of-the-art.
\end{abstract}

\maketitle

\pagestyle{\vldbpagestyle}
\begingroup\small\noindent\raggedright\textbf{PVLDB Reference Format:}\\
\vldbauthors. \vldbtitle. PVLDB, \vldbvolume(\vldbissue): \vldbpages, \vldbyear.\\
\href{https://doi.org/\vldbdoi}{doi:\vldbdoi}
\endgroup
\begingroup
\renewcommand\thefootnote{}\footnote{\noindent
This work is licensed under the Creative Commons BY-NC-ND 4.0 International License. Visit \url{https://creativecommons.org/licenses/by-nc-nd/4.0/} to view a copy of this license. For any use beyond those covered by this license, obtain permission by emailing \href{mailto:info@vldb.org}{info@vldb.org}. Copyright is held by the owner/author(s). Publication rights licensed to the VLDB Endowment. \\
\raggedright Proceedings of the VLDB Endowment, Vol. \vldbvolume, No. \vldbissue\ %
ISSN 2150-8097. \\
\href{https://doi.org/\vldbdoi}{doi:\vldbdoi} \\
}\addtocounter{footnote}{-1}\endgroup

\ifdefempty{\vldbavailabilityurl}{}{
\vspace{.3cm}
\begingroup\small\noindent\raggedright\textbf{PVLDB Artifact Availability:}\\
The source code, data, and/or other artifacts have been made available at \url{\vldbavailabilityurl}.
\endgroup
}

\section{Introduction}
Executing queries in database management systems involves accessing, joining, and aggregating data from various sources. These queries have multiple execution plans, with significant variations in performance. To find the best plan at compile-time, query optimizers use limited search algorithms and a cost model. The cost model evaluates each plan based on parameters such as cardinalities, which are estimated as they are unknown at compile-time. Moreover, the cost model makes certain assumptions that may not hold in practice. Consequently, the plan selected by the optimizer may be suboptimal at runtime, lacking robustness against estimation errors and non-conforming environments. Query optimization approaches that are less sensitive to estimation errors and do not rely on consequential simplifying assumptions are considered robust. This work focuses on the robustness in the context of plan optimization. Robustness in other phases of query optimization and processing, like query rewrite and execution, is beyond the scope of this research.
While various approaches are proposed in the literature for robust query optimization, they are rarely adopted in practice due to their limitations. Some require intrusive changes in the query execution engine \cite{babu_proactive_2005, markl_robust_2004, moumen_handling_2016}, have unreasonable compilation overhead \cite{dutt_plan_2016, karthik_platform-independent_2016, karthik_concave_2018}, or make unrealistic consequential assumptions \cite{christodoulakis_implications_1984,chu_least_2002, wolf_robustness_2018}. With the advent of new approaches based on ML, some of these limitations are mitigated. For example, Neo \cite{marcus_neo_2019} shows some robustness to misestimation of the input parameters, and Bao \cite{marcus_bao_2021} demonstrates minimal regressions for the tested workloads. However, 
ML-based approaches suffer from significant errors when there is a drift in the data or workload distributions \cite{hilprecht_deepdb_2020}. While ML-based techniques show promising advantages for reducing the maintenance and tuning overheads of query optimizers, a successful transition to such approaches relies on addressing concerns about their robustness. 
\textcolor{black}{Given the limitations in prior work, a practical approach to solve the robustness problem needs to a) have limited compilation overheads, b) avoid making consequential or unjustified simplifying assumptions, and c) demonstrate robustness based on well-defined robustness measures.}
We propose Roq, a novel approach for robust query optimization using approximate probabilistic ML. 
Our contributions are: 
\begin{itemize}
    \item We formalize the problem of robust query optimization. 
    \item We formalize the notion of plan and cost model robustness to enable an objective evaluation of these characteristics.
    \item We formalize theoretical methods for quantifying uncertainties and risks in the context of plan evaluation and selection based on approximate probabilistic ML.
    \item We propose a new model architecture for a learned cost model that supports measuring uncertainties and provides improved performance compared with the state-of-the-art. 
    \item We propose novel strategies and algorithms for plan evaluation and selection that outperform the state-of-the-art approaches by accounting for uncertainties and risks.
    \item We demonstrate the robustness of the proposed approaches to workload shifts and their limited compilation overhead.
    \item We conduct an ablation study of the impacts of accounting for different types of risk in the risk-aware strategies.
\end{itemize}

\textcolor{black}{The proposed risk quantification and measurement techniques, along with the risk-aware strategies, are not tightly coupled with the proposed model architecture. Any NN-based learned cost model can benefit from employing these techniques to enhance performance and robustness with minimal overhead. 
}

In the rest of the paper: Section 2 presents the problem of robust query optimization. Section 3 describes the theoretical framework of Roq. Section 4 introduces the architecture of a risk-aware learned cost model. Section 5 presents the experimental study. Section 6 discusses related work, and Section 7 concludes the paper.

\section{Problem Statement}
Query optimization involves exploring a large space of candidate execution plans, recursively evaluating a set of plan fragments (sub-plans), selecting the one with the minimum expected cost, and constructing larger plans from smaller fragments, until the creation of a complete plan. In comparing plan fragments, query optimizers do not consider the uncertainty of cost estimations; i.e., they practically assume that the estimated costs have zero variance. Therefore, they target expected optimality, but not robustness.\footnote{In this paper, when we attribute characteristics to query optimizers, we refer to that of Db2 as a representative of the state-of-the-art commercial RDBMSs.} Although we recognize the differences among the optimization approaches of data systems, to the best of our knowledge, the limitations impacting their robustness are prevalent in the major ones.  \textcolor{black}{In the following, Section \ref{sec:problem_discussion} discusses intuitively the robustness problem and defines types of risk. Section \ref{sec:problem_formulation} provides formal problem definitions using two alternative approaches which are later used in the proposed plan selection strategies in Section \ref{sec:riskaware_strategies}.}  

\subsection{Problem Discussion}
\label{sec:problem_discussion}
Uncertainty in sub-plan evaluation and selection is traced back to three main sources: uncertainty rooted in the (a) plan structure and the data that flow through the plan nodes, (b) limitations of the estimation method, and (c) a crude comparison of a set of plans. We use the terms \textit{risk} and \textit{risky} to point out a high level of \textit{uncertainty} in each of these dimensions. We use the term \textit{robust} if the \textit{sensitivity to uncertainty} is low. Hence, a plan can be considered robust if its execution cost is not sensitive to changes in the parameter estimates and the environment variables. The level of plan robustness is influenced by its structural characteristics, the operators used in the plan, the patterns of error propagation, and the characteristics of the underlying data. Similarly, an estimation method can be considered robust if it is not sensitive to inaccurate parameter estimates and non-conforming environments. Also, an estimation method should quantify the expected uncertainty for each estimate in order to support robustness. Classic cost models are not robust since they are sensitive to inaccurate parameter estimates and non-conforming environments, and do not quantify estimation uncertainties. Due to the uncertainties rooted in the plan structure and data, and the limitations of the cost estimation method, picking a plan from a set of plans is also an uncertain task, as there is a risk that the selected plan will be suboptimal at runtime. We define the sources of uncertainty (or else, risk) in plan evaluation and selection: 

\begin{definition}
    Plan risk is the uncertainty inherent to the plan, influenced by predicates and operators involved in the plan, the plan structure, error propagation patterns, and the data characteristics.
\end{definition}

\begin{definition}
    Estimation risk is the uncertainty in cost estimates due to limited knowledge of the design and parameters of the ideal cost model. It is influenced by limitations in cost modeling, such as simplifying assumptions and error-prone parameter estimates.
\end{definition}

\begin{definition}
    Suboptimality risk is the likelihood of a plan, selected as optimal from a set of plans, being suboptimal at runtime.
\end{definition}

\subsection{Problem Formulation}
    \label{sec:problem_formulation}
We first formulate the classical problem of optimal plan selection; based on this, we provide two alternative formulations of the problem of robust plan selection. The latter are the basis for the proposed risk-aware plan selection strategies presented in Section \ref{sec:riskaware_strategies}.

\textbf{Optimal plan selection problem:} Given a finite set of candidate execution plans $\mathcal{P} = \{p_1, p_2, \ldots, p_n\}$, and a cost function $(f(.))$ that estimates the cost of executing any plan $\{p_i, \forall i \in \{1,\ldots,n\}\}$, find the 
plan $p^*$ such that:
$p^* = \arg\min_{p_i \in \mathcal{P}} f(p_i).\square$  

The above problem formulation ignores that estimated costs are inherently inaccurate and that they may lead to selecting plans that are suboptimal at runtime. In contrast, robust query optimization is defined as "an effort to minimize the \textit{sub-optimality risk} by accepting the fact that estimates could be inaccurate \cite{yin_robust_2015}." Accordingly, we formulate the problem of robust plan selection as follows:

\textbf{Robust plan selection problem (Approach 1):} Given a finite set of candidate execution plans $\mathcal{P} = \{p_1, p_2, \ldots, p_n\}$, a cost function $f(.)$  and a robustness function $g(.)$ that estimate the cost and robustness of any plan $\{p_i, \forall i \in \{1,\ldots,n\}\}$ respectively, find the plan $p^*$ such that:
$ p^* = \arg\min_{p_i \in \mathcal{P}} SOR(p_i)$
where $SOR(p_i)$ is the \textit{suboptimality risk} of the plan $p_i$ compared to the alternatives, defined as a function $k(.,.)$ of $f(.)$ and $g(.)$, i.e. $SOR(p_i) = k(f(p_i),g(p_i))$. $\square$

This approach formalizes the problem as a risk minimization problem, where solving it requires providing approximations for the cost function $\hat{f}(.)$, the robustness function $\hat{g}(.)$ and a formulation for the suboptimality risk function $k(.,.)$. We use this formalization to propose a risk-aware plan selection strategy based on suboptimality risk in Section \ref{sec:subopt_strategy}

Alternatively, we can formalize the problem using an approach that jointly optimizes both cost and robustness. This formalization can be formulated in various ways, such as multi-objective optimization, optimizing a nonlinear combination of the two objectives. Enumerating all such formulations is beyond the scope of this study. Inspired by approaches proposed in the literature \cite{babcock_towards_2005}, we choose to construct a linear combination of cost and robustness functions, as a joint objective function.

\textbf{Robust plan selection problem (Approach 2):} Given a finite set of candidate execution plans $\mathcal{P} = \{p_1, p_2, \ldots, p_n\}$, a cost function $f(.)$  and a robustness function $g(.)$ that estimate the cost and robustness of any plan $\{p_i, \forall i \in \{1,\ldots,n\}\}$ respectively, find the plan $p^*$ such that:
$ p^* = \arg\min_{p_i \in \mathcal{P}} \left(\alpha \cdot f(p_i) + (1 - \alpha) \cdot g(p_i)\right) \square$

Solving the problem based on this formalization necessitates providing an approximation for the cost function $\hat{f}(.)$, the robustness function $\hat{g}(.)$, and a principled approach for determining a suitable weight parameter $\alpha$. 

Most of the prior work has been concerned with providing a better approximation for the cost function $\hat{f}(.)$ \cite{marcus_neo_2019,marcus_bao_2021,chen_loger_2023,zhang_simple_2023, hilprecht_one_2022, siddiqui_cost_2020}. However, providing clear definitions and approximation methods for $\hat{g}(.)$ and $\hat{k}(.,.)$ are missing in the literature. In this paper, we aim at filling in this gap.

\section{Theoretical Framework}

We study the sources of uncertainty in the context of plan evaluation and selection towards formalizing the notion of robustness. \textcolor{black}{Section 3.1 presents a study and definitions for the notions of plan robustness and cost model robustness, and forms a foundation for the design of a robust cost model. Section 3.2 presents quantification measures for the three types of risks tailored for an ML-based cost model. Section 3.3 presents plan evaluation and selection strategies and algorithms that employ risk measures to ensure robustness.}

\subsection{Studying and Modeling Robustness}
\label{sec:robustness}
The uncertainty of a plan cost estimate is rooted in two main sources: a) the plan structure and the data that flow through the plan operators and b) the limitations of the cost model.

\subsubsection{Plan and Cost Estimation Uncertainty Modeling}

    \begin{figure}
        \centering
            \includegraphics[width=1\linewidth]{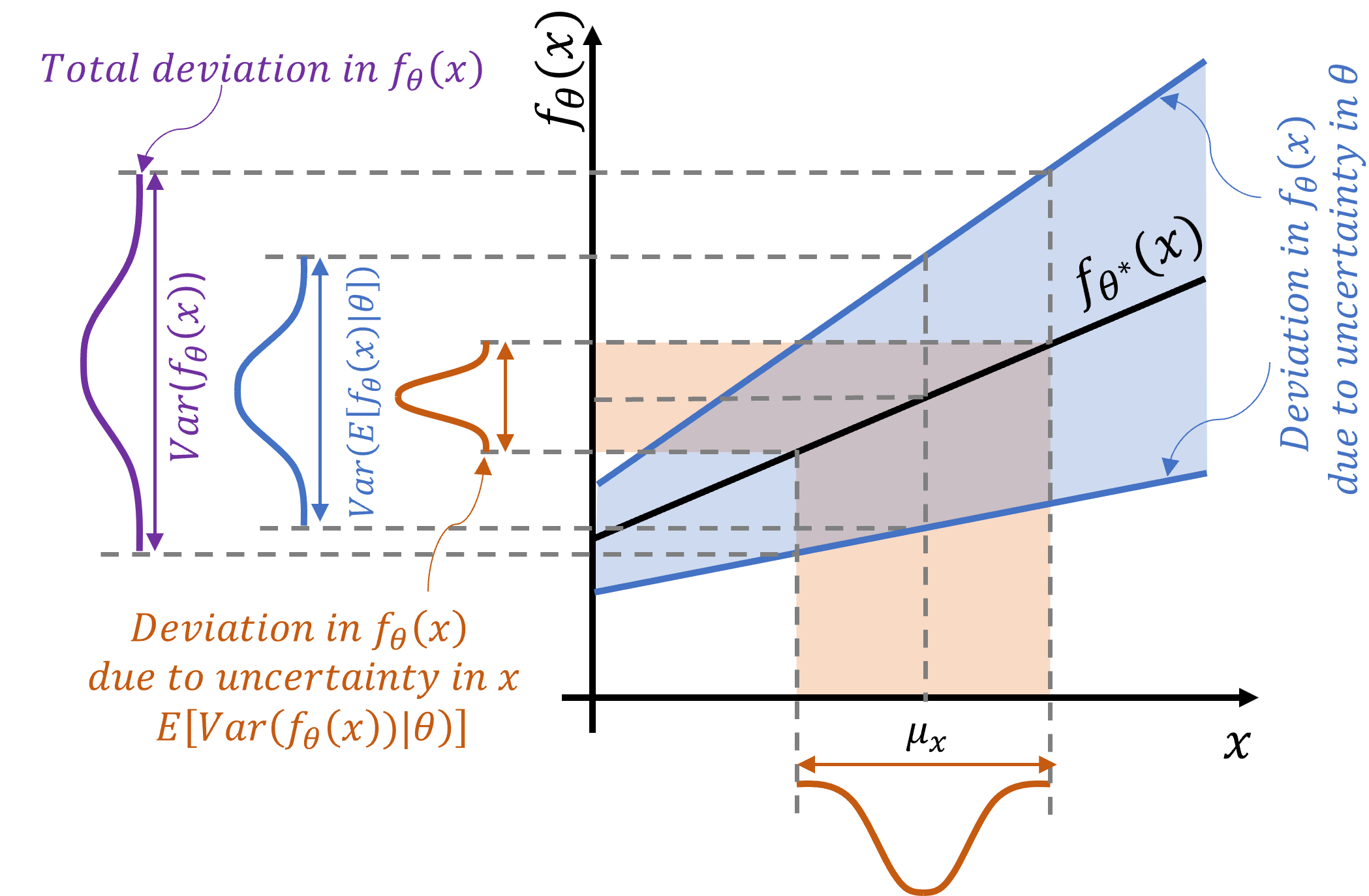}
            \caption{Cost Model Uncertainty Decomposition}
            \label{fig:uncertainty_modeling}
    \end{figure}

Plan robustness can be formalized based on cost behavior given various parameter values~\cite{wolf_robustness_2018}. The plan cost is represented by a Parametric Cost Function (PCF) 
in a multidimensional parameter space, which includes all combinations of values of different error-prone parameters that impact the plan cost. Previous works have suggested various properties of the PCF to measure plan robustness, such as the slope of the PCF and its area under the curve~\cite{wolf_robustness_2018}. 
We formalize this notion by modeling the uncertainty in the PCF and how it can be decomposed into uncertainties from various sources, including the plan structure and the parameters of the cost model.

Consider a PCF $f_\theta(\mathcal{X})$ with error-prone input parameters $\mathcal{X}=\{x_1,...,x_n\}$, and a set of model parameters $\theta$. The uncertainty in the estimation provided by $f_\theta(\mathcal{X})$ can be traced back to the uncertainty in a) input parameter estimates $\mathcal{X}$ and b) model parameters $\theta$. This uncertainty can be decomposed using the law of total variance:\vspace{-0.05in}

\begin{equation}
\label{robustness-decomposition}
\begin{split}
\text{Var} (f_\theta(\mathcal{X})|\mathcal{X}=\mathcal{X}^*) &= E [\text{Var} (f_\theta(\mathcal{X})|\mathcal{X}^*,\theta)] \\
&+ \text{Var} (E[f_\theta(\mathcal{X})|\mathcal{X}^*,\theta])
\end{split}
\end{equation}\vspace{-0.05in}

\noindent where $\mathcal{X}^*$ is the estimated distribution of the input parameters. 


\begin{theorem}
\label{plan-cost-robustness}
The term $E [\text{Var} (f_\theta(\mathcal{X})|\mathcal{X}^*,\theta)]$ represents the uncertainty rooted in the plan structure and its sensitivity to cardinality misestimations and the term $\text{Var} (E[f_\theta(\mathcal{X})|\mathcal{X}^*,\theta])$ represents the uncertainty rooted in model parameters. $\square$
\end{theorem}

The decomposition provided in Theorem \ref{plan-cost-robustness} is illustrated in Figure \ref{fig:uncertainty_modeling}, with a simple linear cost model with one error-prone cardinality parameter. The first term ($E [\text{Var} (f_\theta(\mathcal{X})|\mathcal{X}^*,\theta)]$) has a direct relationship with a) the error in the input cardinality and b) the slope of $f_\theta(\mathcal{X})$ which determines its sensitivity to the error in the input. Therefore, its quantification is a suitable measure for \textit{plan robustness}. The second term ($\text{Var} (E[f_\theta(\mathcal{X})|\mathcal{X}^*,\theta])$), on the other hand, is directly influenced by the uncertainty in model parameters, and is a suitable measure for \textit{model robustness}. This represents the uncertainty rooted in our lack of knowledge about an ideal cost model. A proof for Theorem \ref{plan-cost-robustness} is provided in Appendix \ref{sec:appendixA.1}.

\subsection{Ensuring Robustness via Risk Quantification}
\label{sec:risk_quantification}
We recognized the probabilistic nature of cost estimates and provided a decomposition model for their uncertainties. We continue with the definition of quantification measures of the three types of risks, which will be employed to ensure robustness. The proposed quantification is tailored to an ML-based cost model and employs approximate probabilistic ML. We describe an example with scenarios of plan cost uncertainties that is used in the discussion of the proposed measures.

\begin{figure}[t]
  \centering
  \includegraphics[width=\linewidth]{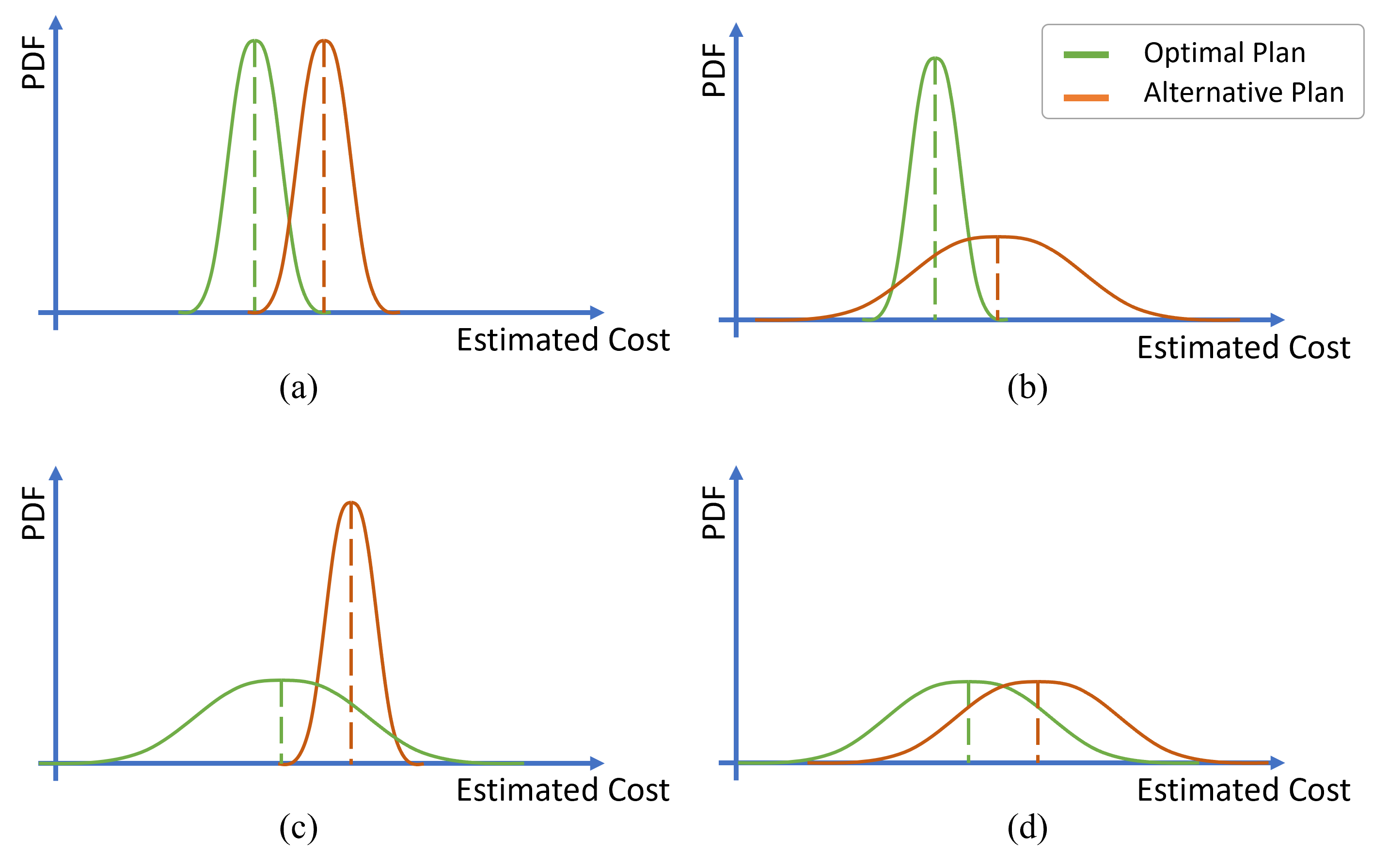}\vspace{-0.1in}
  \caption{Scenarios for plan cost uncertainties for an expected optimal plan and an alternative (second-best) plan.
  }
  \Description{}
  \label{fig:cost_distr_comp}\vspace{-0.1in}
\end{figure}

\begin{example}
\label{ex:cost_distr_comp}
Figure \ref{fig:cost_distr_comp} illustrates various scenarios of cost uncertainties. Each plot shows the probability density function (PDF) of the estimated cost of two alternative plans. The costs are assumed to have a normal distribution. In each case, assuming the optimizer estimates the most likely cost for each plan, the plan with the minimum expected cost is selected as the optimal. The alternative plan, while having a higher expected cost, is also likely to be optimal at runtime with a non-zero probability. Note that the best-case scenario for the optimizer is when the plan costs are consistently accurate, as the optimizer naively assumes (Figure \ref{fig:cost_distr_comp}a). In such a case, the likelihood of plan X being more expensive than plan Y is low. More likely, however, the estimates have various levels of uncertainty. Figure \ref{fig:cost_distr_comp}b/c/d represent such scenarios. If one of the two estimates is relatively more uncertain, the risk can increase substantially (Figure \ref{fig:cost_distr_comp}b/c). The worst case is when both estimates are highly uncertain (Figure \ref{fig:cost_distr_comp}d), as the risk of choosing the expected optimal plan increases further. This example highlights the importance of taking into account the distributions of the cost estimates rather than solely the expected values. $\square$
\end{example}


\subsubsection{Uncertainty Modeling in ML}
The uncertainties captured in Equation \ref{robustness-decomposition} depend on the marginal distribution $p(f_\theta(\mathcal{X})|\mathcal{X}^*,\theta)$. This distribution is conditioned on the cost model parameters and the input parameters. In the context of learned cost models, the model parameters are obtained through training based on labeled samples $\mathcal{D} = \{(x_i,y_i) | \, \forall i \in \{1, \ldots, n\} \}$. Therefore, the distribution must be conditioned on the training samples $\mathcal{D}$:  $p(f_\theta(\mathcal{X})|\mathcal{X}^*,\mathcal{D})$. Making inferences through this distribution comes down to evaluating the following equation and is called Bayesian inference \cite{hullermeier_aleatoric_2021}:
\begin{equation}
    p(f_\theta(\mathcal{X}) | \mathcal{X}^* , \mathcal{D}) = \int p(f_\theta(\mathcal{X}) | \mathcal{X}^* , \theta)p(\theta | \mathcal{D}) \, d\theta
\end{equation}
The marginal distribution $p(\theta | \mathcal{D})$ represents the uncertainty in the parameters of the model given the training samples $\mathcal{D}$. This term captures the additional variance existing in a learned model compared with an analytical (classical) model, and is caused by the stochastic nature of the training process. Therefore, the decomposition of the total variance provided in Equation \ref{robustness-decomposition} can be rewritten as follows:
\begin{equation}
\label{ml-uncertainty-decomposition}
\begin{split}
    \text{Var} (f_\theta(\mathcal{X})|\mathcal{X}=\mathcal{X}^*) 
    &= E [\text{Var} (f_\theta(\mathcal{X})|\mathcal{X}^*,\mathcal{D})] \\
    &\quad + \text{Var} (E[f_\theta(\mathcal{X})|\mathcal{X}^*,\mathcal{D}])
\end{split}
\end{equation}
\vspace{-0.05in}

The first term in Equation \ref{ml-uncertainty-decomposition} is aleatoric (or data) uncertainty and the second term is epistemic (or model) uncertainty \cite{kendall_what_2017}. The variability of the behavior of the real-world phenomena is explained by data uncertainty. Data uncertainty can also be caused by noisy input vectors or labels, as well as low-dimensional input vectors that do not adequately explain the sample \cite{hullermeier_aleatoric_2021}. Model uncertainty captures the variability of predictions due to lack of knowledge about an ideal model. This is rooted in limitations in model architecture, training procedure, or the coverage of the training data \cite{gawlikowski_survey_2021}. For a learned cost model, these notions can be quantified using various techniques from the ML literature \cite{gawlikowski_survey_2021}. In the following, we explain our approach to quantify these two types of uncertainty.

\textbf{Data and Model Uncertainties} A neural network can be designed to predict the parameters of probability distribution \cite{nix_estimating_1994}. i.e. it not only predicts the conditional expected value but also the conditional variance of the target given the training data and the input sample ($E[Var(y|x^*,D)]$ with $y=f_\theta(\mathcal{X})$). This is done by adding a second branch to the output of the neural network, which predicts the variance. Optimizing the following loss function corresponds to maximizing the log-likelihood with a Gaussian prior \cite{nix_estimating_1994}:
\begin{equation}
    \label{eq:loss_function}
    \text{Loss} = \frac{1}{N} \sum_{i=1}^{N} \left( \frac{\ln(\sigma_i^2)}{2} + \frac{(y_i - \mu_i)^2}{2\sigma_i^2} + \frac{1}{2}\ln 2\pi \right) \quad
\end{equation}

\noindent where $N$ is the number of samples in $D$, $\mu_i$ is plugged in from the first branch of the output layer, $\sigma_i^2$ from the second one, and $y_i$ from the labels.

Model uncertainty is captured by the variance of the conditional expected values of the target ($Var(E[(y|x^*,D)])$). 
To capture this, we use approximate variational inference by Monte Carlo (MC) Dropout \cite{gal_dropout_2016}. In this method, the neural network is trained by applying dropout on hidden layers with a Bernoulli distribution. At inference time, $T$ predictions are made with dropout applied randomly to the hidden layers, resulting in variations in the weight matrices, hence in the $T$ predicted values. With a Gaussian prior, this corresponds to combining multiple Gaussian distributions $\{N(\mu_{\hat{\theta}_t},\sigma_{\hat{\theta}_t}^2)\}_{t=1}^T$ into one $N(\hat{\mu},\hat{\sigma}^2)$. Given each draw of model weights results in two estimated parameters, we have a set of parameter estimates $\{(\hat{\mu}_t,\hat{\sigma}_t^2)\}_{t=1}^T$. The predicted mean is:\vspace{-0.1in}

\begin{equation}
    \hat{\mu} = E_{y\sim p(y|x^*,D)}[y] \approx \frac{1}{T} \sum_{t=1}^{T} (\hat{\mu}_t) \quad
\end{equation}\vspace{-0.1in}

The data, model, and total uncertainties are captured by the following equations respectively \cite{kendall_what_2017}:\vspace{-0.1in}
\begin{equation}
\label{eq:data_unc_ml}
    \hat{\sigma}_d^2 = E[Var(y|x^*,D)] = \frac{1}{T} \sum_{t=1}^{T} (\hat{\sigma}_t^2)
\end{equation}\vspace{-0.1in}

\[
\hat{\sigma}_m^2 = \text{Var}\left(E[(y|x^*,D)]\right) = E\left[E\left[(y|x^*,D)\right]^2\right] - \left(E\left[(y|x^*,D)\right]\right)^2 \\
\]\vspace{-0.1in}
\begin{equation}
\label{eq:model_unc_ml}
= \frac{1}{T} \sum_{t=1}^{T} \hat{\mu}_t^2 - \left(\frac{1}{T} \sum_{t=1}^{T} \hat{\mu}_t\right)^2
\end{equation}\vspace{-0.1in}

\begin{equation}
\hat{\sigma}^2 = \hat{\sigma}_d^2 + \hat{\sigma}_m^2
\end{equation}\vspace{-0.1in}

\subsubsection{Risk Quantification}
With the base concepts of uncertainty modeling in ML, in this section the risks associated with plan evaluation and selection are quantified. In addition, the assumptions made and the computation overheads involved are discussed too.

\textbf{Plan Risk} 
As explained in Section \ref{sec:robustness}, data uncertainty captures the uncertainty rooted in input parameters and the sensitivity of the plan to those uncertainties.  Therefore, we relate the riskiness (or robustness) of a plan to data uncertainty and quantify it by the function provided in Equation \ref{eq:data_unc_ml}. Consider the optimal plan in Figure \ref{fig:cost_distr_comp}c. Let us assume the variance of the distributions represents the data uncertainty (i.e., the model uncertainty is zero). Although the expected cost of this plan is lower than that of the alternative plan, its potential deviation at runtime is larger. Running this plan in various conditions and in presence of different parameter values, is likely to lead to longer execution times in some scenarios. The cost for the alternative plan, however, has a tighter distribution while having a higher expected cost. In this example, the parameter representing the deviation of the cost distribution can be considered as the risk inherent to the plan itself. 

\textbf{Estimation Risk} Estimates are inherently subject to errors. Therefore, regardless of how robust a plan is, the estimate of its cost is also subject to uncertainties. 
We use as a measure for the Estimation Risk the model uncertainty as quantified using the function in Equation \ref{eq:model_unc_ml}. 
The scenario represented in Figure \ref{fig:cost_distr_comp}c can be studied assuming that the deviation represents model uncertainty (i.e., data uncertainty is zero). In such a case, the deviation of the cost from the expected value quantifies the uncertainty rooted in the modeling limitations. Therefore, a plan with a wider distribution will have a higher estimation risk. 

\textbf{Suboptimality Risk} 
As stated in the problem formulation with Approach 1, the goal of robust query optimization is to find a plan with minimum risk of suboptimality. This can be formalized as the likelihood of the plan being suboptimal compared with the alternative plan(s). To explain this, let us revisit Example \ref{ex:cost_distr_comp}. We assume that the cost of the optimal plans in all scenarios is 8, and it is 10 for the alternative plans. The difference between the scenarios is the standard deviation of the cost distributions. Let us assume the cost for the optimal plan (plan X) is denoted by $C(X) \sim \mathcal{N}(\mu_x,\sigma_x^2)$ and the cost for the alternative plan (plan Y) is denoted by $C(Y) \sim \mathcal{N}(\mu_y,\sigma_y^2)$. Also, let us assume the covariance between $C(X)$ and $C(Y)$ is denoted as $\text{Cov}(x,y)$. Then the probability of the alternative plan being cheaper than the optimal plan is denoted by:\vspace{-0.1in}

\begin{equation}
    R(p_X, p_Y) = P(C(X) - C(Y) > 0)
\end{equation}\vspace{-0.1in}

\noindent where $R(p_X,p_Y)$ is 
the risk of picking $X$ over $Y$. Note that:
\begin{equation}
D = C(X) - C(Y) \sim \mathcal{N}(\mu_x - \mu_y, \sigma_x^2 + \sigma_y^2 - 2\text{Cov}(x,y))
\end{equation}

For the sake of simplicity, let us assume $X$ and $Y$ are independent variables, therefore $\text{Cov}(x,y) = 0$. Then:
\begin{equation}
\label{eq:SOR_dist}
D = C(X) - C(Y) \sim \mathcal{N}(\mu_x - \mu_y, \sigma_x^2 + \sigma_y^2)
\end{equation}

The probability of $C(X) - C(Y) > 0$ is then obtained by computing the z-score for $C(X) - C(Y) = 0$:\vspace{-0.1in}
\begin{equation}
\text{zscore} \approx \frac{0 - (\mu_x - \mu_y)}{\sqrt{\sigma_x^2 + \sigma_y^2}} = \frac{\mu_y - \mu_x}{\sqrt{\sigma_x^2 + \sigma_y^2}}
\end{equation}\vspace{-0.1in}

In all 4 scenarios, $\mu_x = 8$, $\mu_y = 10$, and $\mu_y - \mu_x = 2$. Table \ref{tab:risk_computation} outlines the computation of the risk associated with choosing the optimal plan for all scenarios illustrated in Figure \ref{fig:cost_distr_comp}.

\begin{table}[t]
\centering
\caption{Computation of risks associated with selecting the optimal plan at different uncertainty levels}
\label{tab:risk_computation}
\begin{tabular}{cccccc}
\hline
\textbf{Scenario} & $\boldsymbol{\sigma_x}$ & $\boldsymbol{\sigma_y}$ & $\boldsymbol{\sqrt{\sigma_x^2+\sigma_y^2}}$ & \textbf{z-score} & \textbf{P(C(X)-C(Y)>0)} \\ \hline
a                 & 1                       & 1                       & $\sqrt{2}$                                         & $\approx 1.41$    & $\approx 7.9\%$          \\
b                 & 1                       & 4                       & $\sqrt{17}$                                        & $\approx 0.49$    & $\approx 31.6\%$         \\
c                 & 4                       & 1                       & $\sqrt{17}$                                        & $\approx 0.49$    & $\approx 31.6\%$         \\
d                 & 4                       & 4                       & $\sqrt{32}$                                        & $\approx 0.35$    & $\approx 36.3\%$         \\ \hline
\end{tabular}
\end{table}

Note that the computed risks are consistent with the qualitative study of the graphs in Figure \ref{fig:cost_distr_comp}. These computations can be generalized to compute the risk for choosing the optimal plan in the presence of several alternative plans to give the Suboptimality Risk. This can be computed as the average risk of picking the target plan over any other plan in the search space, as follows:\vspace{-0.1in}

\begin{equation}
\label{eq:SOR}
    \text{SOR}(p_i) = \frac{1}{S-1} \sum_{j=1, j \neq i}^{S} R(p_i, p_j)
\end{equation}\vspace{-0.1in}

\noindent where $S$ is the total number of plans in the search space, SOR$(p_i)$ is Suboptimality Risk for plan $p_i$ in the presence of every other $S-1$ plan, and $R(p_i, p_j)$ is the risk of picking the plan $p_i$ over plan $p_j$. Note that this risk factor can be computed using either model or data uncertainty for $\sigma$ values. The former gives the risk rooted in modeling limitations while the latter gives the risk rooted in the plan structure. In Roq, we choose to compute this value using the total uncertainty to account for both types of risk. We demonstrate in Appendix \ref{sec:appendixA.2} that $\text{SOR}(p_i)$ is a function of the expected cost and robustness of all plans in the search space.

\textbf{Computation Overheads} In order to minimize the compilation overheads, the computation of SOR can be vectorized, such that at each step of comparing several plan fragments $R(p_i,p_j)$ is computed as a matrix $R_{n \times n}$ where $n$ is the number of plans to be compared and $R_{i,j}=R(p_i,p_j)$. To this end, we define $D_{n \times n}$ where $D_{i,j}=\mu_i - \mu_j$ and $S_{n \times n}$ where $S_{i,j}=\sqrt{\sigma_i^2 + \sigma_j^2}$. Then the z-scores for comparing different pairs of plans can be computed as:  $Z_{n \times n} = -{D_{n \times n}}\oslash{S_{n \times n}} \quad
$, where the $\oslash$ symbol denotes the Hadamard (elementwise) division. The computation overheads using this vectorization approach are negligible as demonstrated in the experimental study.

\textbf{The Independence Assumption} Let us now revisit Equation 9 without assuming independence. Note that the covariance component can be computed as:

    $\text{Cov}(x,y) = E(C(X) \cdot C(Y)) - E(C(X)).E(C(Y)) \quad $

An accurate estimation of the risk associated with selecting the plan $X$ over the plan $Y$ would require the probability distribution of the cost for each plan, as well as the pairwise joint distribution $C(X) \cdot C(Y)$.  Assuming independence between $C(X)$ and $C(Y)$ results in overestimating the risk by increasing the variance of $C(X) - C(Y)$. We assume independence for two reasons: a) the computation is impractical and the overhead is prohibitive, and b) even with the independence assumption made, the risk quantification provides significant improvements. The effectiveness of the simplified quantification, as demonstrated in the experimental study, can be explained by the fact that the overestimation effect for all pairs of plans is consistent, effectively canceling each other out.

\begin{figure}
    \centering
    \includegraphics[width=1\linewidth]{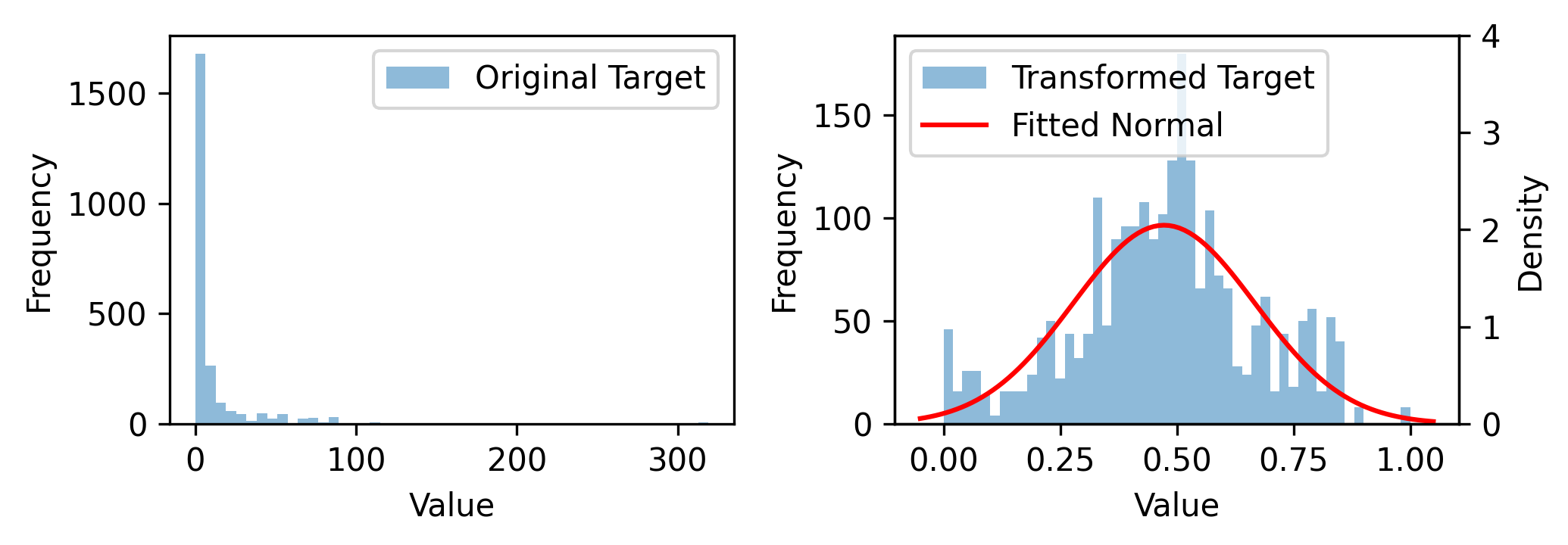}
    \caption{\textcolor{black}{Target distributions before and after transformation}}
    \label{fig:target_transformation}
\end{figure}

\textbf{The Probability Distribution} The cost and execution time are not inherently normally distributed variables, as assumed by the loss function (Equation \ref{eq:loss_function}). However, the targets can be transformed so that they approximately conform to a normal distribution. This is handled in the preprocessing phase as explained in Section \ref{Experimental Setup}. \textcolor{black}{The impact of the transformation on the target values is shown in Figure\ref{fig:target_transformation}, where the latency distributions of the JOB benchmark before and after the preprocessing step are shown.}

\textcolor{black}{The above discussions support that the assumptions made in Roq are both justified and inconsequential as stated in the design objective (b) in the introduction.} 

\begin{algorithm}[t]
\caption{Plan Selection by SubOpt Risk ($\mathcal{P}, \mathcal{C}$)}
\label{algo:subopt_risk}
\begin{algorithmic}[1]
\STATE \textbf{Inputs:} \\
    $\mathcal{P} = \{ p_i \, | \, \forall i \in \{1, \ldots, S\} \}$ Set of plans to be evaluated \\
    $\mathcal{C} = \{ C_{p_i} \sim N(\mu_{p_i}, \sigma_{p_i}^2) \, | \, \forall i \in \{1, \ldots, S\} \} $ cost distributions \\
\STATE \textbf{Output:} Selected plan $P^*$
\FOR{$p_i \in \mathcal{P}$}
    \STATE $\Delta_{ij} \sim N(\mu_i - \mu_j, \sigma_i^2 + \sigma_j^2)$ for all $j \neq i$
    \STATE $R(p_i) = \frac{1}{S-1} \sum_{j \neq i} P(\Delta_{ij} > 0)$
\ENDFOR
\STATE $P^* = \arg \min_{p_i \in \mathcal{P}} R(p_i)$
\STATE \textbf{return} $P^*$
\end{algorithmic}
\end{algorithm}


\begin{algorithm}[t]
\caption{Conservative Plan Selection ($\mathcal{P}, \mathcal{C}, f_s$)}
\label{algo:conservative_plan_selection}
\begin{algorithmic}[1]
\STATE \textbf{Inputs:} \\
    $\mathcal{P} = \{ p_i \, | \, \forall i \in \{1, \ldots, S\} \}$ Set of plans to be evaluated \\
    $\mathcal{C} = \{ C_{p_i} \sim N(\mu_{p_i}, \sigma_{p_i}^2) \, | \, \forall i \in \{1, \ldots, S\} \} $ cost distributions \\
    $f_s$: a value greater than 0 \\
\STATE \textbf{Output:} Selected plan $P^*$
\STATE $P^* = \arg \min_{p_i \in \mathcal{P}} (\mu_{p_i} + f_s \cdot \sigma_{p_i})$
\STATE \textbf{return} $P^*$
\end{algorithmic}
\end{algorithm}

\subsection{Risk-aware Plan Evaluation}
    \label{sec:riskaware_strategies}
Classic optimizers select a plan from a set of plans solely based on expected cost. They assume that the estimated expected costs have zero uncertainties. In contrast, when the uncertainties, and, thus, risk, are quantified, a new family of plan evaluation and selection methods can be devised that account for them. We propose three risk-aware plan evaluation strategies. While the first two use risk measures directly for selecting plans, the third uses risk to prune the search space. We design algorithms that implement these strategies.

\subsubsection{Plan Selection by Suboptimality Risk}
    \label{sec:subopt_strategy}
This strategy follows the first approach to the problem formulation where the problem is defined as a risk minimization one. The suboptimality risk is computed for every plan in the search space, and the one with the minimum risk is selected, based on the risk quantification method suggested in Section \ref{sec:risk_quantification}.
Given a set $\mathcal{P} = \{p_i \, | \, \forall i \in \{1, \ldots, S\}\}$ of $S$ plans, it computes the suboptimality risk $R(p_i)$ for every $p_i$. The plan with a minimum $R(p_i)$ is selected. The algorithm can use either plan or estimation risk, or a combination of the two.

\subsubsection{Conservative Plan Selection}
    \label{sec:conservative_strategy}
This strategy assumes that each plan has a higher execution cost than its estimated cost, and this difference is proportional to the standard deviation of its estimated cost. It eliminates risky plans (i.e., plans with high standard deviation) that happen to be low cost. Given a set $\mathcal{P} = \{p_i \, | \, \forall i \in \{1, \ldots, S\}\}$ of $S$ plans, where each $p_i$ has a cost distribution $C_{p_i} \sim N(\mu_{p_i}, \sigma_{p_i}^2)$, rather than picking the plan with minimum $\mu_{p_i}$, it picks the plan with minimum $\mu_{p_i} + f_s \cdot \sigma_{p_i}$, where $f_s$ is a parameter tuned on a validation set. The algorithm can be specialized to use either the plan or the estimation risk, or both. It follows the second approach to the problem formulation where a linear combination of the two objectives $\alpha \cdot \mu_{p_i} + (1-\alpha) \cdot \sigma_{p_i}$ is minimized, where $f_s = \frac{1 - \alpha}{\alpha}$.

\subsubsection{Search Space Pruning by Plan and Estimation Risks}

This strategy prunes the search space to eliminate highly risky plans, based on the plan or the estimation risks. Algorithm~\ref{alg:pruning} implements this strategy: given a set $\mathcal{P} = \{p_i \mid \forall i \in \{1, \dots, S\}\}$ of $S$ plans along with their plan and estimation risks ($R_p$ and $R_e$ respectively), and fractions of plans to be pruned with the highest plan and estimation risks ($f_{pr}$ and $f_{er}$ respectively), thresholds $\varphi_{pr}$ and $\varphi_{er}$ are computed for plan and estimation risks, respectively. Any plan with either plan or estimation risks above the thresholds is eliminated. Parameters $f_{pr}$ and $f_{er}$ are tuned using the samples in the validation set. Choosing the plan from the remaining ones relies on a plan selection strategy.



\begin{algorithm}[t]
\caption{Pruning by plan and estimation risks $(\mathcal{P}, R_e, R_p, f_{er}, f_{pr})$}
\label{alg:pruning}
\begin{algorithmic}[1]
\STATE \textbf{Inputs:} \\
  $\mathcal{P} = \{p_i \mid \forall i \in \{1, \dots, S\}\}$: a set of plans to be evaluated \\
  $R_e = \{R_e(p_i) \mid \forall i \in \{1, \dots, S\}\}$: a set of estimation risks \\
  $R_p = \{R_p(p_i) \mid \forall i \in \{1, \dots, S\}\}$: a set of plan risks \\
  $f_{er}$: a fraction of plans to be pruned by estimation risk \\
  $f_{pr}$: a fraction of plans to be pruned by plan risk
\STATE \textbf{Output:} $P' \subset \mathcal{P}$: pruned search space
\STATE $\varphi_{er} = \text{sorted}(R_e)[\lfloor |\mathcal{P}| \times f_{er} \rfloor]$
\STATE $\varphi_{pr} = \text{sorted}(R_p)[\lfloor |\mathcal{P}| \times f_{pr} \rfloor]$
\STATE $P' = \{p_i \in \mathcal{P} \mid R_e(p_i) \leq \varphi_{er} \text{ and } R_p(p_i) \leq \varphi_{pr}\}$
\STATE \textbf{return} $P'$
\end{algorithmic}
\end{algorithm}

\begin{figure*}[t]
  \centering\vspace{-0.05in}
  \includegraphics[width=1\linewidth]{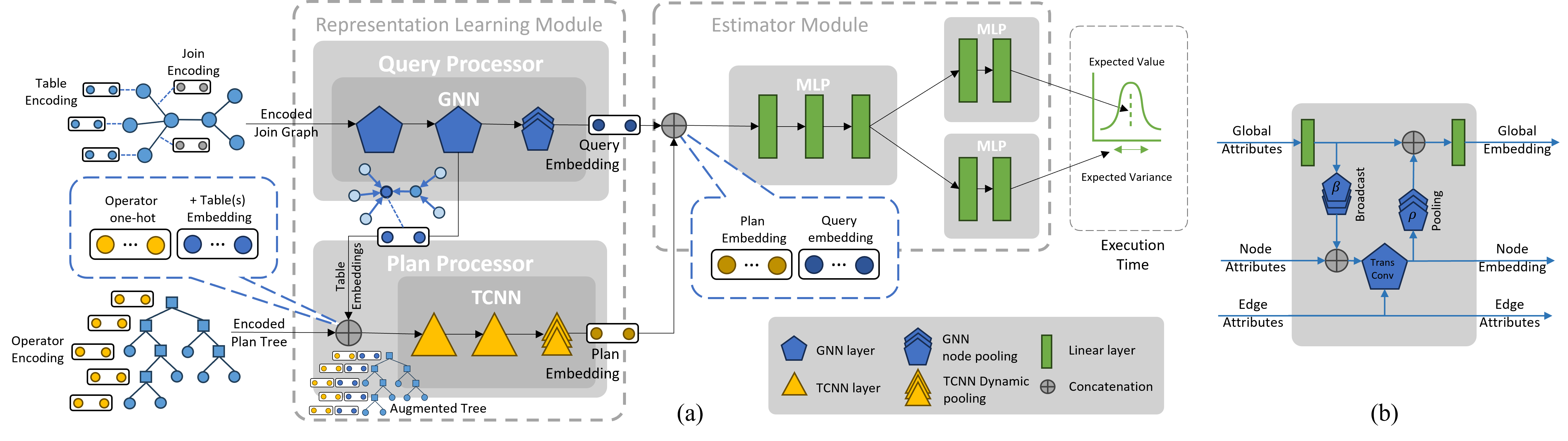}
  \caption{a) Architecture of the risk-aware learned cost model, including representation learning and estimator components. This architecture allows quantification of model and data uncertainties, \textcolor{black}{b) Architecture of the extended transformerConv GNN model block that enables receiving and processing graph level attributes in addition to node and edge attributes}}
  \Description{}
  \label{fig:Roq}
\end{figure*}

\section{Risk-aware Learned Cost Model}

We propose a robust learned cost model, 
which infers the execution time, and the plan, estimation, and suboptimality risks. Figure \ref{fig:Roq}-a illustrates the model's architecture. It takes as input queries and plans and predicts the execution time along with expected variance. It 
includes two main modules: The first learns the representation of query and plan combinations; the second takes this representation and learns the query-plan associations 
and their impact on runtime.

\textbf{Query and Plan Encoding.}
The model employs the following encoding of input queries and plans.

\noindent\textit{Query Encoding.} 
We use the query’s join graph to encode tables, local predicates, join predicates, aggregations, etc, similarly to previous work \cite{yu_cost-based_2022}. In this work, we encode table statistics, such as cardinality and selectivity from local predicates 
and correlations between predicate columns (captured by the optimizer statistics) as node attributes. We incorporate characteristics of joins, such as join type, join predicate operator, join column skewness and join selectivity, as edge attributes. The optimizer statistics are extended to capture the join characteristics, too. In addition to node and edge attributes, we capture high-level characteristics of the join graph, such as the number of tables and joins, and the join graph topology, as global attributes. This representation is agnostic to the plan (join orders and plan operators) that will be used to execute the query. 

\noindent\textit{Plan Encoding.} We use a vectorized plan tree that preserves the structure of the plan \cite{marcus_neo_2019}. Each node corresponds to a plan operator and captures the operator type, including different access, join, and aggregate operators, using one-hot encoding. In addition, it captures the table(s) in the subgraph originating at the node using an embedding learned by the Query Processor (see Figure \ref{fig:Roq}-a). 

\textbf{Representation Learning.}
The representation learning module 
produces query embeddings (Query Processor) and plan embeddings (Plan Processor). Although we partially borrow the architecture of the Plan Processor from prior work \cite{marcus_neo_2019}, the architecture of the Query Processor and the output that feeds the Plan Processor are novel and support the robustness of the proposed model.

\noindent\textit{Query Processor.} 
We use Graph Neural Nets (GNNs) to process the encoded join graph and learn query and table representations. Each GNN layer serves as a message passing step in which each table receives information from its adjacent tables, each additional layer adding to the depth of the message passing in the graph. This allows learning a representation for each table such that it contains information about its adjacent tables and the whole join graph. Additionally, we use mean and max pooling to aggregate the information from all tables and produce a representation for the whole join graph. Although there are various GNN implementations, we use TransformerConv \cite{shi_masked_2021} in our architecture. The TransformerConv uses an attention mechanism \cite{vaswani_attention_2017} to condition the amount of information each node receives from its neighboring nodes by the attributes of the target node, the neighboring nodes, and the edges connecting them to the target node. This is a suitable option for processing the join graphs as the amount of data processed through a subgraph should be conditioned by the tables and the joins involved. 
\textcolor{black}{We extend TransformerConv as illustrated in Figure \ref{fig:Roq}-b, to receive and process graph-level attributes, broadcast them to every node, and apply mean and max pooling to aggregate them back into a graph level embedding. This novel architecture allows us to effectively produce query embeddings.}

\noindent\textit{Plan Processor.} 
The Plan Processor takes the vectorized plan tree as input and augments nodes with the embedding of the corresponding table(s) produced by the Query Processor. 
If a node involves multiple tables, their corresponding embeddings are aggregated using mean-pooling before being concatenated to the plan node. The augmented plan tree is then processed through multiple layers of Tree Convolutional Neural Networks (TCNN), ultimately aggregated into a one-dimensional vector using dynamic pooling \cite{mou_convolutional_2016}. This architecture is proven to be effective in processing query plan trees by prior works \cite{marcus_neo_2019,marcus_bao_2021,yu_cost-based_2022}. In our tests, TCNN was superior to alternative architectures for processing trees such as GNNs or Tree-LSTMs \cite{yu_rtos_2020}. The embeddings generated by the final layer of this model serve as the representation of the query plan. 

\textbf{Estimation Module.}
The Estimation Module takes the query and plan representations as input. These two are concatenated into a single vector which is then processed through a multi-layered perceptron (MLP) module. This module learns the association between the query and plan representations. The output of this module is then fed to two separate branches of MLPs. The first branch predicts the expected execution time, while the second branch predicts the expected variance (uncertainty) of the execution time.

\textcolor{black}{
\textbf{Novelty of Roq's Model Architecture. }
While some components of Roq's model are derived from prior works, to the best of our knowledge, their composition and extension in this architecture towards learning cost and quantifying risks are novel. Specifically, the novelty lies in the adoption and extension of the state-of-the-art GNNs to learn generalizable representations for tables and join graphs and their integration into TCNNs, as well as the employment of two separate branches of MLPs for predicting the plan cost and risk simultaneously. 
}\vspace{-0.1in}

\section{Experimental Study}

We experimentally compare Roq with the state-of-the-art.\vspace{-0.05in}

\subsection{Experimental Setup}\label{Experimental Setup}


\subsubsection{Database and environment settings.}\label{Database and environment settings} 
We use {IBM Db2} for compiling and running the queries. Plans obtained from the {Db2} optimizer and the best-performing plans are used as baselines for evaluating the plans selected by the proposed approaches. We consider the plan with the minimum execution time among all plans in the search space as the best performing plan.\vspace{-0.05in} 

\subsubsection{Datasets and workloads.}
We use the following benchmarks: The Join Order Benchmark (JOB) \cite{leis_how_good_2015}, the Cardinality Estimation Benchmark (CEB) \cite{flowloss}, and the Decision Support Benchmark (DSB) \cite{nambiar_making_2006}. 
The characteristics of the workloads used are summarized in Table \ref{tab:workload}. Training data includes pairs of queries and plans, 
encoded to be consumable by an ML model, as described in the following.

\textbf{\textit{CEB}} 1000 queries are randomly sampled from query templates in CEB with up to 12 joins over 21 tables from the IMDB dataset. These are split to non-overlapping subsets of 800, 100, and 100 queries for training, validation, and test. Each experiment is repeated 5 times with different random seeds and the average results are reported.

\textbf{\textit{JOB}} contains 113 queries with up to 27 joins over 21 tables from the IMDB dataset. Given the small sample size, 10-folds cross validation is used to train models using this dataset. In each fold, 80\% are used for training, 10\% for validation, and 10\% for testing, such that over the 10 folds each query appears in the test set exactly once. The aggregated test results from different folds are reported.

\textcolor{black}{\textbf{\textit{DSB}} 1000 queries are generated based on the Decision Support Benchmark (DSB) \cite{ding_dsb_2021}, an extension of TPC-DS that enhances data generation by introducing more realistic, skewed, and correlated data distributions rather than the mostly independent, uniform distributions in TPC-DS. It augments query templates with predicate filters for fine-grained data slicing and adds new templates featuring more complex join patterns (such as non-equi and many-to-many joins), resulting in many more distinct query instances. The queries are generated based on 15 distinct query templates and are split into 800, 100, and 100 queries for training, validation, and test sets, respectively. Each experiment is repeated 5 times with different random seeds and the average results are reported.}

\textbf{Plan Generation} Each query is compiled with various hint sets and the different plans generated by the optimizer are captured. The first compilation is done without any constraints, leading to the use of the default plan chosen by the optimizer. Additionally, each query is compiled using 13 hint sets (see Appendix \ref{hint-sets}). 
Hint sets produce locally optimal plans that are expected to be better than the optimizer's plan in some scenarios, as shown by Bao \cite{marcus_bao_2021}. Although this does not represent the way the optimizer traverses the search space, it is sufficient for providing a diversified set of potentially optimal plans for the purpose of this study.

\begin{table}[t]
\centering
\caption{Benchmarks used in the experiments}\vspace{-0.1in}
\label{tab:workload}
\begin{tabular}{cccccc}
\hline
\textbf{ } & \textbf{Queries} & \textbf{Templates}   & \textbf{Max Joins} & \textbf{Dataset} & \textbf{Tables} \\ \hline
CEB     & 1000      & 16          & 12          & IMDB           & 21         \\ 
JOB     & 113       & 33          & 28          & IMDB           & 21         \\ 
DSB   & 1000	    & 15          & 13           & TPC-DS      & 24         \\ 
\hline
\end{tabular}
\end{table}

\textbf{Collecting Labels} Each query is executed using each of the guidelines and the execution time is measured.
\textcolor{black}{A timeout threshold is used to terminate too long executions. 
This threshold is dynamically set to be 10 times the time it takes to run the query using the best performing plan found so far, rounded up to the nearest greater integer. This threshold allows a higher level of exploration to obtain information on more risky plans.}

\textbf{Preprocessing} Preprocessing is performed for all splits based on the statistics obtained from the training data. Min-max scaling is applied to bring the range of the values for each feature between zero and one:
    $X_t = \frac{{X - \min(X)}}{{\max(X) - \min(X)}}$.
%
We apply logarithmic transformation and min-max scaling to labels. The execution times can range greatly and typically exhibit a large skewness towards zero. Log transformation is applied to the labels to reduce skewness. This is performed with the following formula, where \( y_{log} \) represents the transformed label:
   $ y_{log} = \log_{10}(y)$.
%
The Min-Max scaling then brings the range of values between zero and one. This is a desirable range for deep learning models that use the Sigmoid activation function in their output layers:
    $y_t = \frac{{y_{log} - \min(y_{log})}}{{\max(y_{log}) - \min(y_{log})}}$. 


\subsubsection{Model Training and Parameter Tuning.} To avoid over-fitting, three mechanisms are used in training: early stopping, dropout, and reducing learning rate on plateau. Applying dropout at every layer is necessary to enable the variational inference needed to obtain model uncertainty. The dropout rate along with other parameters such as the number of layers, number of neurons, and learning rate are tuned using the Asynchronous Successive Halving Algorithm \cite{li_system_2020}. This allows for exploring a large number of parameter combinations while limiting the total time needed for tuning.

\subsubsection{Baselines} 
\textcolor{black}{We compare Roq against a classical RDBMS optimizer, that of {IBM Db2}, as well as four prominent works, Neo \cite{marcus_neo_2019}, Bao \cite{marcus_bao_2021}, Lero \cite{zhu_lero_2023}, and Balsa \cite{yang_balsa_2022}. Neo and Bao are selected due to their demonstrated robustness to errors in input features. Balsa is selected due its to demonstrated robustness to workload shifts. Lero is selected due to its demonstrated minimization of regressions. Each one 
is evaluated for its ability to select a robust plan from the same set of plans for a given query. To this end, the search space is identical for all approaches and is enumerated as explained.}

\textbf{\textit{{IBM Db2}}} serves as a baseline representing the state-of-art classical query optimizers in our experiments. In our tests, Db2’s optimizer captures Column Group Statistics (CGS) to account for correlations between database attributes. This feature reduces the impact of the independence assumption which leads to severe cardinality and cost underestimations. The performance and robustness of the plans selected by Db2 serve as a baseline.

\textbf{\textit{Neo}} is a fully learned query optimizer that uses deep neural networks and reinforcement learning to generate efficient query execution plans. It starts by bootstrapping from traditional optimizers and then continuously refines its decision-making by learning from actual query execution times \footnote{\textcolor{black}{Our implementation of Neo uses one-hot encoding for table representation. In order to provide it with the knowledge of the underlying data, we add cardinality estimates to encodings for each plan node.}}.

\textbf{\textit{Bao}} uses a more lightweight cost model that takes plans as input and predicts latency. At each node of the vectorized plan tree, in addition to node type, it captures the estimated cardinality and cost.

\textcolor{black}{
\textbf{\textit{Balsa}} is a learned query optimizer that avoids relying on expert demonstrations. It bootstraps from a simple simulator and then fine-tunes its deep reinforcement learning model using real query execution latency feedback.
}

\textcolor{black}{
\textbf{\textit{Lero}} is a more recent work that proposes a learning-to-rank approach to a learned query optimizer. It learns to compare pairs of plans in the search space and create a total order. The model is trained using a pairwise ranking loss function. Additionally, it enumerates pairs of subplans traversed during search and uses them as additional samples to train the model.
}\vspace{-0.05in}

\subsubsection{Evaluation Measures} 
We design five sets of experiments to evaluate Roq based on prediction error, correlations with runtime, robustness to workload shifts, the inference overheads, and an ablation study on the role of plan and estimation risks. 

\textbf{\textit{Prediction Error.}} The predictive performance of the model is evaluated using various measures depending on the subject under consideration. The accuracy of the predicted execution times is evaluated by the Q-error \cite{moerkotte_preventing_2009}. Given the actual label $y$ and the estimated value \( \hat{y} \), Q-error is defined as:
    $Q_{error} = \frac{{\max(y,\hat{y})}}{{\min(y,\hat{y})}} $.

\textbf{\textit{Correlation.}} We measure the correlations between the estimates (or the optimizer’s Cost) and the actual execution times using Spearman’s rank coefficient rather than the more widely used Pearson’s coefficient. While Pearson’s coefficient measures the linear relationships between two variables, Spearman’s coefficient measures the monotonic relationships, whether linear or not. Therefore, given the estimated costs or execution times are used to rank the plans, Spearman’s coefficient is a more suitable measure of correlation.

\textbf{\textit{Plan Suboptimality.}} The quality of plans is evaluated using the suboptimality measure, suggested for end-to-end robustness \cite{haritsa_robust_2020}. The suboptimality of a plan \( p_i \) is computed with respect to the best plan \( p* \):
$\text{Subopt}(p_i) = \frac{ET(p_i)}{ET(p^*)}\in \mathbb{R};  1 < \text{Subopt}(p_i) < \infty$, where $ET(p_i)$ is the execution time of the plan $p_i$, and $p* = argmin_{i \in \{1, \ldots, S\}}ET(p_i)$.
The tail-end of the distribution of suboptimality over all plans, selected by a method, characterizes the worst-case performance degradation of that method compared with an oracle that always selects the best plan. 

\textbf{\textit{Runtime.}} Improving the tail-end suboptimality as a measure of robustness may lead to picking plans that are more robust but more expensive than the optimal plan. While the main goal of our work is to improve robustness, we need to demonstrate that robust query optimization does not cause a major slowdown for a given workload. Therefore, the total runtime of the workload is measured for each approach.

\textbf{\textit{Inference Overheads.}} Using the risk-aware plan evaluation algorithms has an overhead for the inference time since multiple predictions must be made using the base model. We measure the inference time for these algorithms for various inference iterations while evaluating the impact on plan suboptimality and runtime.

\subsection{Experimental Results}
\begin{figure}
  \centering
  \includegraphics[width=1\linewidth]{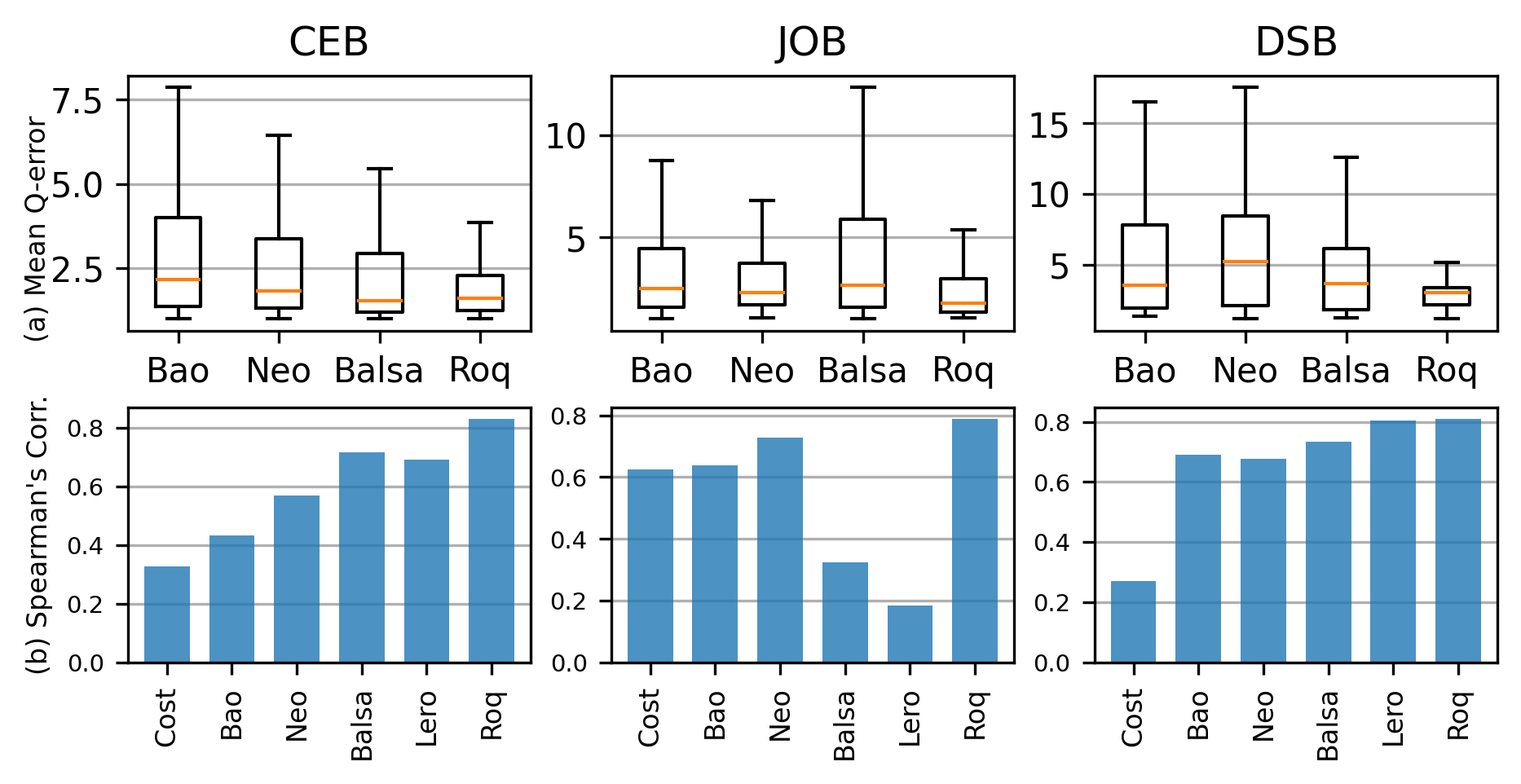}\vspace{-0.15in}
  \caption{\textcolor{black}{Roq's predictive performance vs. the baselines. (a) q-error, and (b) Spearman's correlation with respect to latency. The \textit{Cost} baseline refers to the optimizer's cost estimate}}\vspace{-0.15in}
  \Description{}
  \label{fig:accuracy}
\end{figure}

\textbf{Prediction Accuracy. }
The average predictive performance of Roq's base model (without uncertainty quantification) consistently outperforms the baselines. Roq's prediction's q-error is superior to the baselines in all tested scenarios based on CEB, JOB, and DSB, as shown in Figure \ref{fig:accuracy}a. \footnote{\textcolor{black}{Note that Lero is not included in this evaluation as it is a learning-to-rank model, and that, similar to the optimizer's cost, the absolute values of its predictions do not correspond to latency. Therefore, its q-error is not meaningful.}}
The correlation of the predicted runtime with the actual runtime is more significant for Roq compared with either the baseline ML-based approaches or the Db2 optimizer’s cost \textcolor{black}{(denoted as \textit{Cost})}, as shown in Figure \ref{fig:accuracy}b. In these evaluations we only use the raw predictions from each model and do not use MC dropout variational inference to obtain a more accurate prediction.


\begin{figure*}
  \centering
  \begin{subfigure}{0.49\textwidth}
    \includegraphics[width=\linewidth]{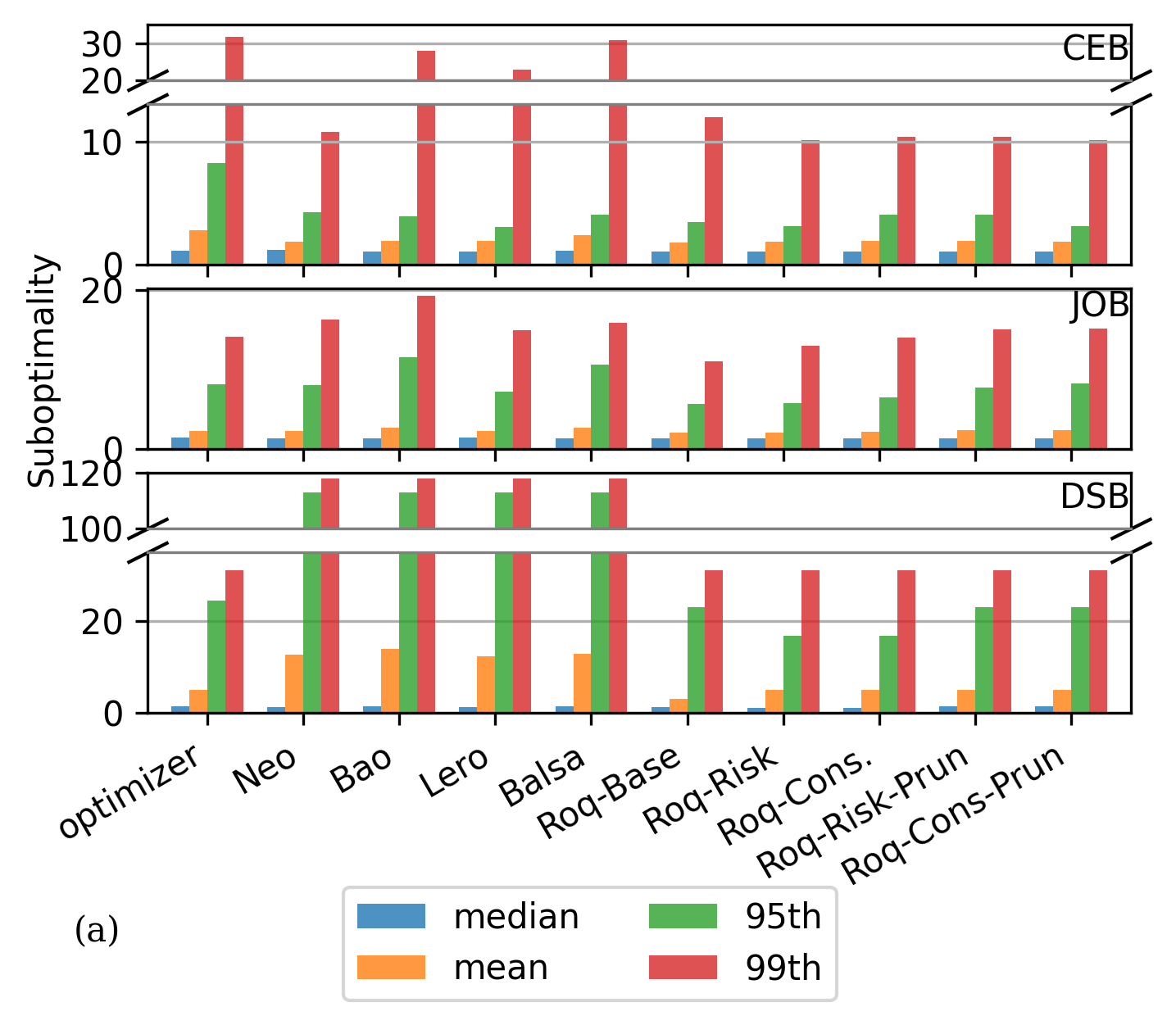}
  \end{subfigure}
  \hspace{0.0cm}
  \begin{subfigure}{0.48\textwidth}
    \includegraphics[width=\linewidth]{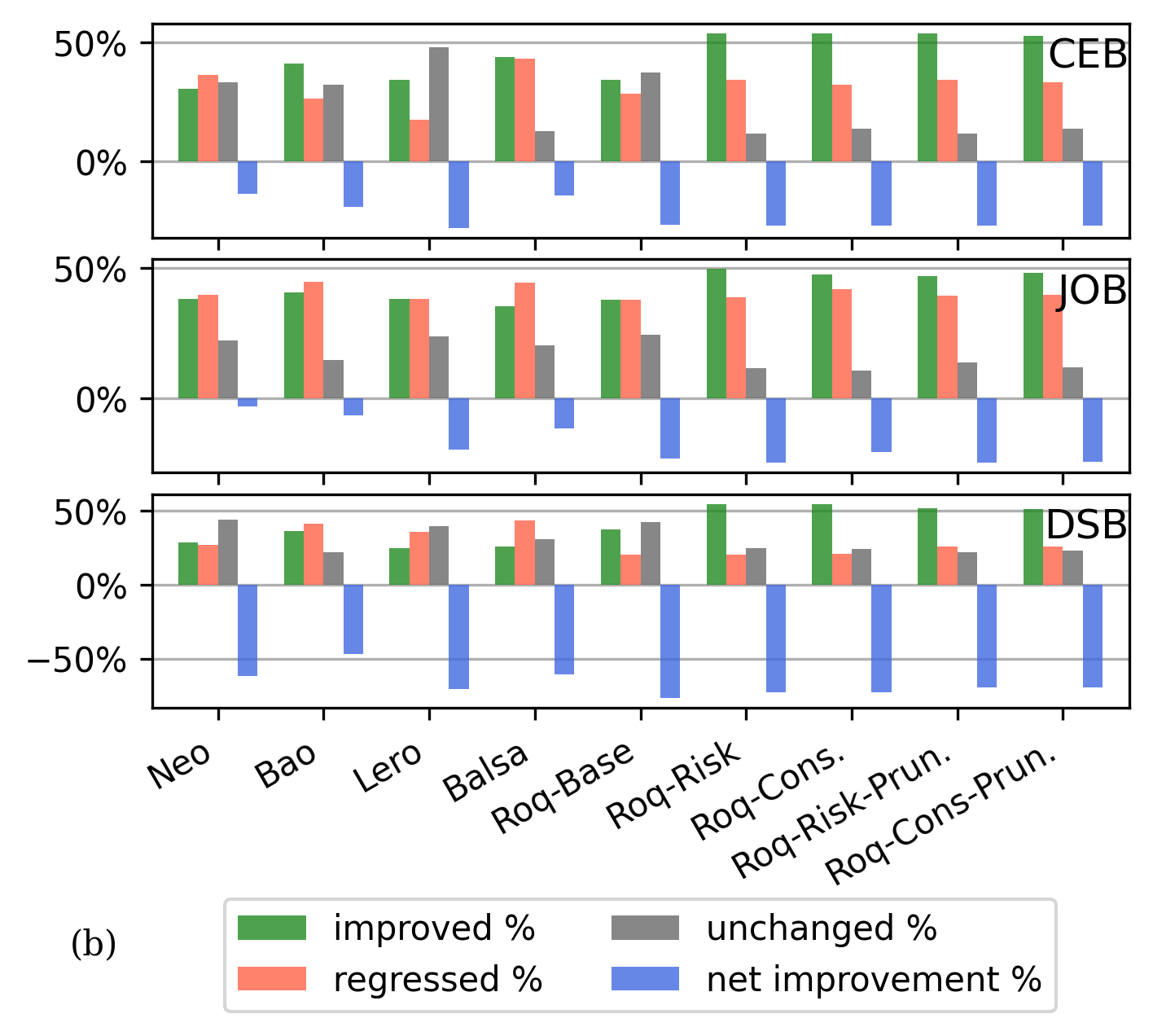}
  \end{subfigure}\vspace{-0.15in}
  \caption{\textcolor{black}{The performance of plans selected by each approach illustrated by (a) suboptimality distribution and (b) improved, regressed, and unchanged percentage of queries, as well as total runtime improvement percentage compared with Db2's plan 
  }}
  \label{fig:subopt_runtime_main}
\end{figure*}

\textbf{Model Architecture. } 
Better average predictive accuracy may not necessarily lead to selecting a better plan. Therefore, we perform experiments to evaluate the quality of the generated plans based on their suboptimality and runtime. The plans selected by the risk-aware strategies using Roq demonstrate greater robustness based on tail-end plan suboptimality and increased runtime improvements compared with the baselines, as shown in Figures \ref{fig:subopt_runtime_main}a and \ref{fig:subopt_runtime_main}b, respectively.  

\textcolor{black}{To isolate the impact of the plan selection strategies as implemented by Algorithms \ref{algo:subopt_risk}, \ref{algo:conservative_plan_selection}, and \ref{alg:pruning}, Roq is tested once without risk quantification, similar to the baselines. This scenario is denoted as \textit{Roq-Base}. On average, Roq’s base model improves the mean and the $99^{th}$ percentile of suboptimality compared with the optimizer by 26.9\% and 28.0\%, respectively. In addition, it improves the same measures compared with the ML-based baselines on average by 30.8.3\% and 44.9.8\% respectively, and it improves runtime compared with the ML-based baselines on average by 22.5\%.
These improvements are attributed to Roq's architecture using GNNs for producing node embeddings from the join graph, which enables enhanced 
generalization to the unseen test queries.}

\textcolor{black}{
\textbf{Plan Selection Strategies.} The risk-aware plan selection strategies effectively increase the number of improved queries and improve the overall performance and robustness. As demonstrated in Figure \ref{fig:subopt_runtime_main}, in comparison with the ML-based baselines, plan selection by subopt risk (denoted by \textit{Roq-Risk}) yields the greatest level of robustness by improving the mean and the $99^{th}$ percentile of suboptimality on average by 24.5\% and 44.4\% respectively. Conservative plan selection (denoted by \textit{Roq-Cons}) improves the same measures by 22.5\% and 44.4\% respectively. These two strategies improve runtime compared with the ML-based baseline on average by 23.0\% and 17.3\% respectively. \textit{Roq-Risk} and \textit{Roq-Cons} improve the $99^{th}$ percentile of suboptimality compared with \textit{Roq-Base} by 19.6\% and 19.4\%, respectively.
While both \textit{Roq-Risk} and \textit{Roq-Cons} improve the tail-end performance similarly, the former was more effective at improving the overall runtime.
These two strategies are combined with plan pruning based on plan and estimation risks, (denoted by \textit{Risk-Roq-Prun} and \textit{Risk-Cons-Prun}). Pruning did not have a significant impact on the performance of either \textit{Roq-Risk} or \textit{Roq-Cons} over the three benchmarks. This is expected as these strategies adjust the predictions according to their uncertainties and effectively eliminate risky predictions by default.}

\begin{table*}[ht!]
\caption{\textcolor{black}{Comparing the impacts of the risk-based plan selection strategies using model, data, and total uncertainties in suboptimaility distribution for CEB, JOB, and DSB queries.}}
\label{tab:subopt_ablation}
\centering
\resizebox{\textwidth}{!}{
  \begin{tabular}{lrrrr rrrr rrrr}
  \toprule
  & \multicolumn{4}{c}{CEB} & \multicolumn{4}{c}{JOB} & \multicolumn{4}{c}{DSB}\\
  \cmidrule(lr){2-5}\cmidrule(lr){6-9}\cmidrule(lr){10-13}
  \textbf{Suboptimality} & \textbf{median} & \textbf{mean} & \textbf{95th} & \textbf{99th}
                        & \textbf{median} & \textbf{mean} & \textbf{95th} & \textbf{99th}
                        & \textbf{median} & \textbf{mean} & \textbf{95th} & \textbf{99th}\\
  \midrule
  base model 
  & 1.05 & 1.16 & 1.44 & 2.68
  & 1.18 & 1.26 & 1.86 & 2.87
  & 6.33 & 5.85 & 13.35 & 16.58 \\
  \midrule
  
  conservative model uncertainty
  & 1.14 & 1.16 & 1.59 & 1.85
  & \underline{\textbf{1.17}} & 1.20 & 1.72 & 1.94
  & 5.49 & 6.02 & 12.87 & \underline{\textbf{14.75}} \\
  
  conservative data uncertainty
  & \underline{\textbf{1.10}} & \underline{\textbf{1.13}} & \underline{\textbf{1.27}} & \underline{\textbf{1.80}}
  & \underline{\textbf{1.17}} & \underline{\textbf{1.18}} & \underline{\textbf{1.56}} & \underline{\textbf{1.83}}
  & 8.44 & 7.05 & 13.94 & 17.10 \\
  
  conservative total uncertainty
  & \underline{\textbf{1.10}} & \underline{\textbf{1.13}} & \underline{\textbf{1.27}} & \underline{\textbf{1.80}}
  & \underline{\textbf{1.17}} & \underline{\textbf{1.18}} & \underline{\textbf{1.56}} & \underline{\textbf{1.84}}
  & \underline{\textbf{4.60}} & \underline{\textbf{5.10}} & \underline{\textbf{11.96}} & 14.96 \\
  \midrule
  
  risk model uncertainty
  & 1.14 & 1.16 & 1.47 & 1.85
  & \underline{\textbf{1.17}} & 1.21 & 1.72 & \underline{\textbf{1.94}}
  & 4.32 & 5.72 & 12.03 & \underline{\textbf{14.08}} \\
  
  risk data uncertainty
  & \underline{\textbf{1.09}} & \underline{\textbf{1.13}} & \underline{\textbf{1.44}} & \underline{\textbf{1.80}}
  & \underline{\textbf{1.17}} & 1.21 & 1.72 & \underline{\textbf{1.94}}
  & 6.50 & 6.60 & 12.97 & 15.28 \\
  
  risk total uncertainty
  & 1.14 & 1.16 & 1.47 & 1.85
  & \underline{\textbf{1.17}} & \underline{\textbf{1.19}} & \underline{\textbf{1.64}} & \underline{\textbf{1.94}}
  & \underline{\textbf{3.76}} & \underline{\textbf{5.36}} & \underline{\textbf{11.15}} & 14.17 \\
  \bottomrule
  \end{tabular}
}
\end{table*}

\textbf{Plan vs. Estimation Risks.} To evaluate the roles of the plan risk (quantified by data uncertainty) and estimation risks (quantified by model uncertainty) on plan selection, we run an ablation study in which the risk-aware plan selection strategies are tested using either model, data, or total uncertainties. \textcolor{black}{The quality of the resulting plans is evaluated based on mean, median, 95th and 99th percentiles of the suboptimality of the resulting plans for all three tested benchmarks (Table \ref{tab:subopt_ablation}).} Using either model or data uncertainty for quantifying risks 
provides significant improvements compared with the base model and using total uncertainties provides maximum improvements. This suggests that the plan and estimation risks each play a role in improving Roq’s performance and the impact is maximized if they are used together. In addition, the impact of accounting for plan risk is more significant for achieving better runtime performance since it effectively avoids long-running plans and thus significantly reduces regressions.

\textbf{Robustness to Workload Shifts.} A major challenge of using ML-based techniques in query optimization is dealing with the problem of generalization to out-of-distribution test samples. 
A learned model becomes obsolete if the characteristics
of the test query differ significantly from the ones used in training. While model retraining is possible, it takes time, and therefore the optimizer would typically need to still rely on the obsolete model in the meantime.
We evaluate the impact of a workload shift for Roq in comparison to the baselines using two simulations.

\textcolor{black}{
In a \textit{\textbf{minor workload shift}} simulation, the baseline model is trained using all query templates in the DSB benchmark. The baseline model is tested on making predictions for two selected templates \textit{query025} and \textit{query101}, due to their unique characteristics
\footnote{\textcolor{black}{\textit{query025} is a 7-join query with a complex join graph involving three aliases of the \textit{`DATE\_DIM`} table, each used for fine-grained temporal slicing of a different fact table with overlapping time windows. It includes many-to-many joins between the fact tables \textit{STORE\_SALES} and \textit{STORE\_RETURNS} using multiple join predicates. \textit{query101} is a query with 9 joins, involving a self-join over the \textit{DATE\_DIM} table and many-to-many joins between the fact tables \textit{STORE\_SALES}, \textit{STORE\_RETURNS}, and \textit{WEB\_SALES}.}}.
The second model is trained using the other 13 query templates in the DSB and tested on the two selected templates.
}

\begin{figure}[t]
    \centering
    \includegraphics[width=1\linewidth]{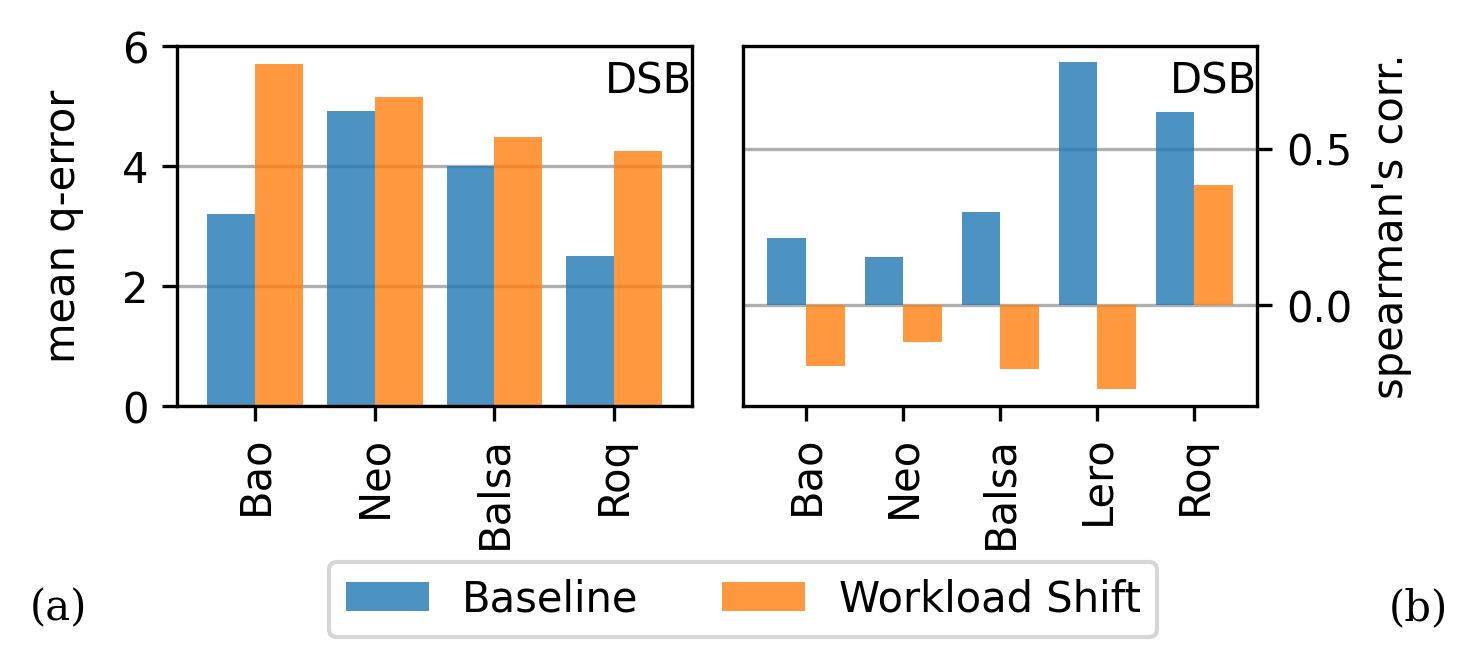}
    \caption{Roq’s generalization to out-of-distribution samples compared with the baselines. a) average q-error, b) Spearman's correlation with runtime}
    \label{fig:dsb_wlsh}
\end{figure}

\begin{figure}[t]
  \centering
  \includegraphics[width=1\linewidth]{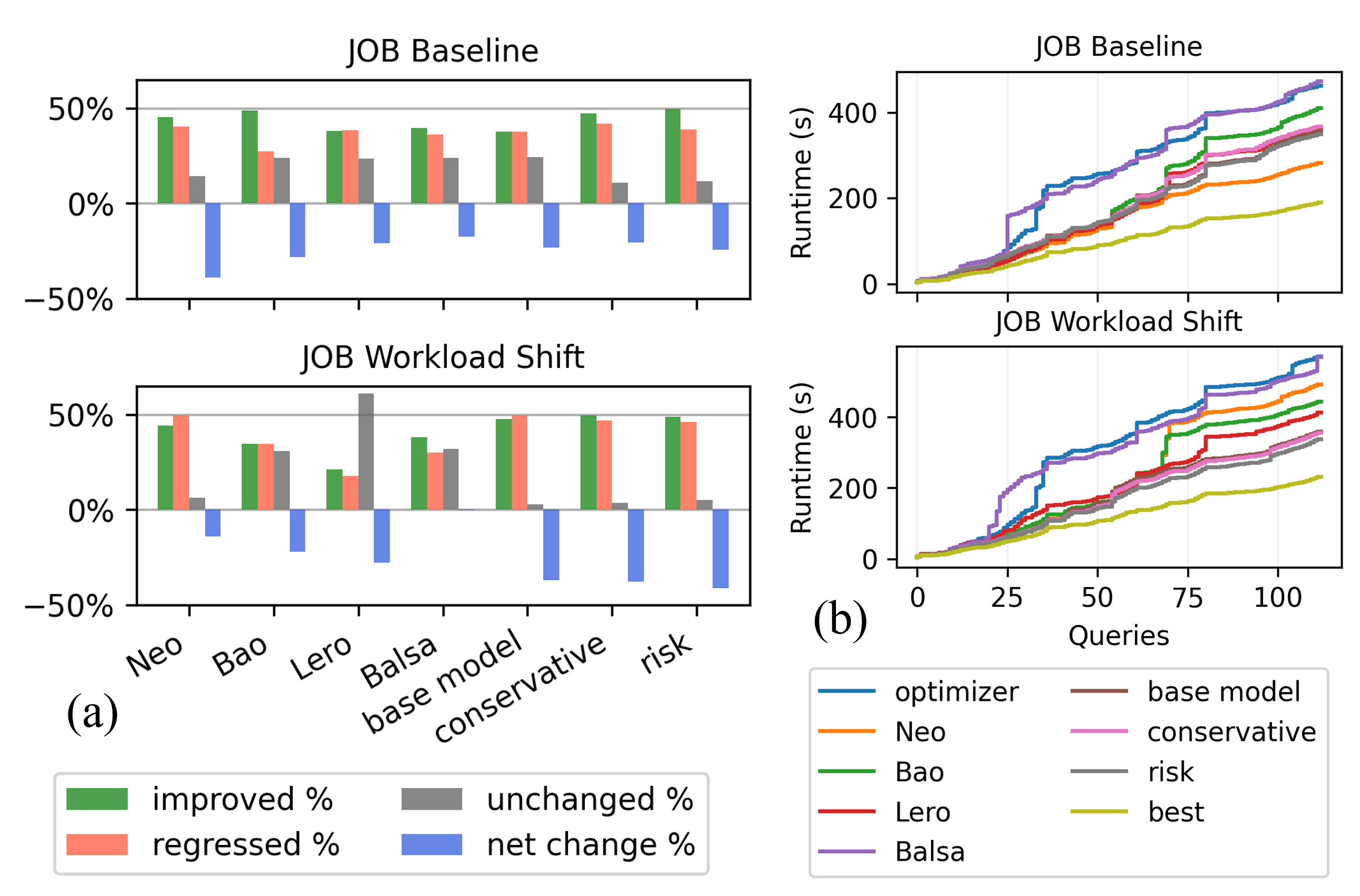}
  \caption{JOB runtime changes after a severe workload shift demonstrated by (a) runtime improvements vs. regressions (b) workload cumulative runtime}
  \Description{}
  \label{fig:job_wlshift}
\end{figure}

A \textit{\textbf{major workload shift}} scenario is simulated using the CEB and JOB. The baseline model is trained and tested on JOB. \textcolor{black}{This represents an ideal scenario where the test workload is identically distributed as the training workload.} The model representing a workload shift is trained on CEB and tested on JOB. This constitutes a major workload shift as the query templates in CEB are entirely different than those in JOB. In addition, our CEB queries have up to 12 joins while JOB has up to 28 joins. 

The results shown in Figure \ref{fig:dsb_wlsh} demonstrate Roq’s robustness to out-of-distribution data caused by a \textit{minor workload shift} in the first scenario. Figures \ref{fig:dsb_wlsh}a and \ref{fig:dsb_wlsh}b show that while the baseline models’ q-error and correlation degrade significantly with a minor workload shift, Roq’s predictive performance degradation is minimal. These results confirm a greater robustness of Roq to out-of-distribution data caused by workload shifts.
\textcolor{black}
{We also evaluate the impact of workload shift on runtime in the \textit{major workload shift} scenario. As demonstrated in Figures \ref{fig:job_wlshift}a and \ref{fig:job_wlshift}b, the performance of three out of four baseline models degrades significantly. Lero's performance slightly improves, which can be attributed to the fact that it does not need to make accurate predictions as long as it can properly rank the plans. Roq's base model's performance also improves compared to the baselines. This can be attributed to the use of GNNs for learning generalizable table and query representations from the CEB workload and use that effectively to make accurate predictions on JOB queries and the fact that our CEB workload includes 1000 queries compared to JOB that includes 113. The additional samples enable a greater performance on an unseen test set when generalizable representations are learned. The Conservative and the Risk strategies, further reduce regressions leading to greater net improvements in runtime. Therefore, Roq and its risk-aware strategies meet the design objective (c) of demonstrating robustness based on well-defined measures.}

\begin{figure*}[h!]
  \centering
   \includegraphics[width=0.89\linewidth]{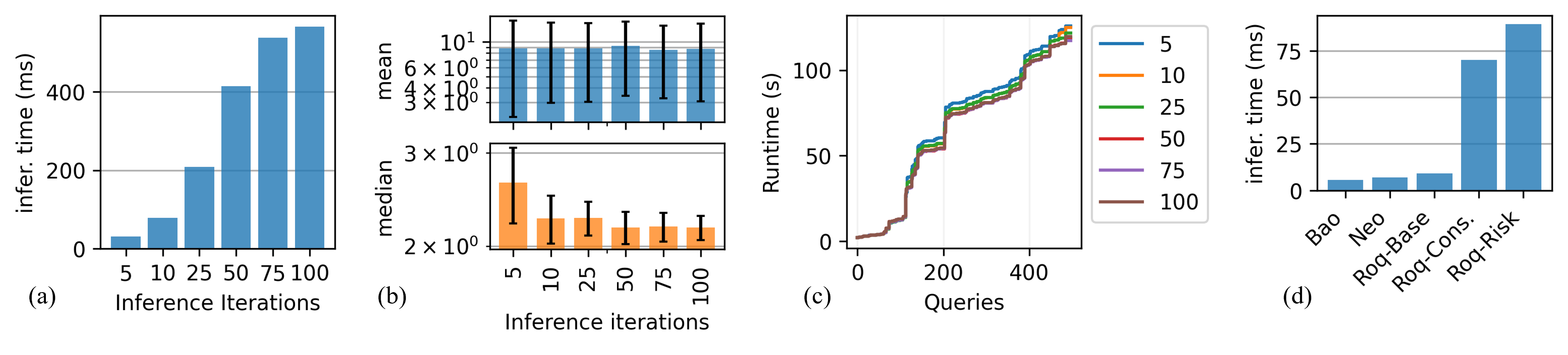}\vspace{-0.15in}
  \caption{\textcolor{black}{The impact of the inference iterations on (a) the inference time, (b) suboptimality, (c) the accumulated runtime of the selected plans and (d) inference overheads for risk-based plan selection methods with 10 inference iterations vs. the baselines}}\vspace{-0.1in}
  \Description{}
  \label{fig:inf_overheads}
\end{figure*}

\textbf{Inference Overheads. }
Capturing estimation risk using the MC Dropout method requires multiple inferences. A higher number of inferences should provide a more accurate estimate of the expected values and the uncertainties. However, this incurs additional overhead for the query compilation phase, which is on the critical path of query execution and must be kept at a minimum. We experiment with inference iterations ranging from 5 to 100. We use plan selection by suboptimality risk as it is expected to have a larger overhead compared to the conservative plan selection strategy, given the more complex computations. We evaluate the inference time, the runtime, and the suboptimality of selected plans. We report an average over 10 runs for each experiment.
Inference times are captured using a single thread on the CPU, typically available for query compilation. As shown in Figure \ref{fig:inf_overheads}a, they grow substantially with the number of iterations. 
Figures \ref{fig:inf_overheads}b and \ref{fig:inf_overheads}c illustrate the impact of increasing the number of inference iterations on the suboptimality and the runtime of the selected plans. Increasing the number of inferences does not have a major impact on mean suboptimality. The median suboptimality has a sharp decline from 5 to 10 inferences but does not show a significant improvement with larger values. The changes in plan runtimes are not significant either. This means the accuracy provided by 10 inferences should be sufficient for selecting robust plans. This demonstrates that our approach is robust to making a limited number of inferences. Figure \ref{fig:inf_overheads}d demonstrates the inference overheads of the risk-aware strategies compared with the baselines {\footnote{\textcolor{black}{Balsa and Lero use Neo and Bao's value networks respectively. Therefore, their inference overheads are identical to Neo and Bao. While Lero is a learning to rank model, it does not require multiple inferences to create a total order \cite{zhu_lero_2023}.}}}. \textcolor{black}{This time measurement does not include plan generation, as this is identical for all alternatives. Overheads for \textit{Roq-Cons} and \textit{Roq-Risk} are around 70 and 89 milliseconds, respectively. As expected, plan selection by suboptimality risk is slightly more expensive than doing so by the conservative strategy. Note that these measurements are performed based on a prototype code in Python and can be significantly reduced with more optimized implementations. Therefore, Roq and its risk-aware strategies meet the design objective (a) of limiting the compilation overheads at a practical level.}

\textbf{Overview of Findings. }
Roq produces predictions with a higher level of accuracy and correlations with runtime. 
The risk quantification methods and plan selection strategies have significant benefits in selecting robust plans. Our experiments show the role of accounting for plan or estimation risks in the plan selection strategies. The proposed strategies demonstrate significant robustness to shifts in workload characteristics. The quality of the resulting plans is robust to a limited number of inference iterations used to quantify risks. Therefore, the proposed techniques can be used with minimal inference overheads. \textcolor{black}{Our observations suggest that \textit{Roq-Risk} and \textit{Roq-Cons} demonstrate similar results. However, from our experience, \textit{Roq-Risk} is preferable as it is non-parametric if the computation overheads can be tolerated. Also, pruning by plan or estimation risk does not provide significant improvements for these strategies. Lastly, our ablation study suggests that accounting for \textit{Plan Risk} is more beneficial than accounting for \textit{Estimation Risk}.}

\section{Related Work}
Query optimization robustness has been studied with a focus on one of the following: re-optimization, parameter value discovery, robustness quantification, adaptive processing, and employing ML.

\textbf{Re-optimization Techniques} use intervals instead of point estimates to identify robust plans within parameter ranges \cite{babu_proactive_2005, markl_robust_2004, moumen_handling_2016}. 
These plans are used at run-time, while actual parameter values or intervals are discovered either via execution or sampling. When values are outside of the robustness range of the plan, the optimizer suggests another plan expected to be robust in the new range. The plan execution terminates prematurely and switches to the new plan. These approaches necessitate intrusive modifications to extend the plans and the execution engine to allow for re-optimization.

\textbf{Discovery-based Techniques} eliminate reliance on parameter estimations by exploring the entire parameter space during compilation. They use mechanisms to iteratively execute plans within time budgets, discarding results when a plan exceeds its allocated time \cite{dutt_plan_2016, karthik_platform-independent_2016, karthik_concave_2018}. 
These methods suffer from significant compilation overheads, making them suitable primarily for OLAP-style environments. Also, they assume a perfect correlation between the optimizer's cost function and the execution time, and they require partial execution of suboptimal plans to discover parameter values.

\textbf{Robustness Quantification Techniques} incorporate robustness alongside optimality in plan selection. One work uses \textit{"Least Expected Cost"} instead of \textit{"Least Specific Cost"}, deriving a PDF of plan costs \cite{chu_least_2002}. This enables selection based on the expected cost across the parameter space rather than a specific parameter setting. 
Deriving the distribution of cost analytically, however, is not a trivial task given the large number of error-prone parameters involved. 
Other works propose a framework to estimate an upper bound for cost given cardinality estimation errors \cite{kastrati_optimization_2016, moerkotte_preventing_2009}. While this can quantify plan robustness, it depends on prior knowledge of the cardinality errors. 
Another work evaluates robustness based on the derivative and area under the curve of the PCF \cite{wolf_robustness_2018}. This approach overlooks the probabilistic nature of the estimated parameters and assumes they are uniformly distributed.
\cite{hertzschuch_small_2021}. In addition, it relies on classical cost models built by making simplifying assumptions 
\cite{christodoulakis_implications_1984}, potentially leading to poor performance even with accurate parameter estimates. Moreover, this approach assumes the PCF is a linear function, which does not hold in practice.

\textbf{Adaptive Techniques} delay some optimization decisions to runtime, when true statistics are obtained. Adaptive Join Algorithm (AJA) \cite{sack_introducing_2019, zhang_simple_2023} switches a hash join to a nested-loops join after determining the true cardinality of the built hash table. 
Look-ahead Information Passing (LIP) \cite{zhu_looking_2017} generalizes a semijoin optimization to a pipeline of equijoins, typically observed in left-deep plans based on star schema queries. Smooth Scan \cite{borovica-gajic_smooth_2018} proposes a new access operator that morphs between a sequential and an index scan depending on statistics at runtime.
These approaches reduce sensitivity to cardinality misestimations 
\cite{zhang_simple_2023}. They are complementary to learning-based techniques, as they improve query processing at runtime while the latter improve query optimization. 

\textbf{Learning-based Techniques} aim to learn the execution cost from runtime \cite{hilprecht_one_2022, marcus_neo_2019, marcus_bao_2021, siddiqui_cost_2020}. These works employ learning from a set of queries and corresponding plans using supervised or reinforcement learning. While one study shows some robustness with respect to estimation errors \cite{marcus_neo_2019}, these works still face uncontrolled predictive uncertainties inherent in learning-based methods. 
They do not address the issue of potential performance regressions compared to classic optimizers. 
Hybrid Cost Model \cite{ wang_hybrid_2023} proposes a meta-model that learns to route queries to suitable cost models (either learned or classic ones) when they are expected to produce better plans, but it does not account for uncertainties or show robustness to out-of-distribution data. A similar study \cite{yu_cost-based_2022} proposes a hybrid approach that combines learning-based cost models with traditional optimizers, selecting partial left-deep join orders based on the predictions of the model. It uses uncertainty estimates to eliminate highly uncertain plans, with a fallback to the classic optimizer when all plans are highly uncertain. 
It does not demonstrate the robustness of produced plans or robustness to out-of-distribution samples. Also, it does not acknowledge that a highly uncertain plan may still be substantially cheaper than a robust plan and that pruning may not be beneficial. In comparison, Roq’s risk-aware strategies use uncertainties not only for pruning but also for plan evaluation and selection by employing suboptimality risk quantification or conservative plan selection strategies. In addition, they allow one to prioritize robustness over optimality (or vice versa). Also, they demonstrate strong robustness not only for unseen queries but also for out-of-distribution samples.\vspace{-0.1in}

\section{Conclusion}

We present Roq, a novel approach based on approximate probabilistic ML that enhances the robustness of query optimization in RDBMSs. We establish a theoretical framework to formalize the concepts of robustness and risk for plan evaluation and selection, including techniques for quantifying plan and estimation risks. We propose risk-aware strategies for plan evaluation and selection that leverage these risk measures, along with a learned cost model incorporating GNN-based query and plan embeddings. In comprehensive experiments, Roq demonstrates superior predictive accuracy, improved robustness to workload shifts, and significant performance gains over state-of-the-art approaches, while maintaining practical compilation overheads.



\balance
\bibliographystyle{ACM-Reference-Format}
\bibliography{references}

\begin{appendices}

  \section{Appendices}
  \subsection{Proof of Theorem \ref{plan-cost-robustness}}
  \label{sec:appendixA.1}
    \begin{proof}
    Let us consider a cost model that is linear with respect to the size of a single error-prone cardinality (Figure \ref{fig:uncertainty_modeling}):
    
    \begin{equation}
        f_\theta(x) = ax + b; \quad \theta=\{a,b\}\\ 
        \label{eq1}
    \end{equation}
    
    where $a$ and $b$ have normal errors: $a \sim \mathcal{N}(\mu_a,\sigma_a^2)$, and $b \sim \mathcal{N}(\mu_b,\sigma_b^2)$, and $x=x^* \sim \mathcal{N}(\mu_{x^*},\sigma_{x^*}^2)$ is the approximated cardinality distribution for the input plan. The expected value $E[f_\theta(x)|x=x^*]$ can be computed a follows:
    
    \begin{equation}
        \begin{aligned}
            E[f_\theta(x)|x=x^*] &= E[a.x^* + b] = E[a.x^*] + E[b] \\
            &= E[a].E[x^*] + E[b] \text{  (a and x are independent)}\\
            &= \mu_a . \mu_{x^*} + \mu_b
        \end{aligned}
    \end{equation}

    The total variance $\text{Var} (f_\theta(x)|x=x^*)$ can be decomposed using the law of total uncertainty:
    
    \begin{equation}
        \text{Var} (f_\theta(x)|x=x^*) = E [\text{Var} (f_\theta(x)|x^*,\theta)] + \text{Var} (E[f_\theta(x)|x^*,\theta])
    \end{equation}
    
    The first and the second components can be computed as follows:

    \begin{equation}
    \label{eq:data_unc}
        \begin{aligned}
            E [\text{Var} (f_\theta(x)|x^*,\theta)] &= E [\text{Var}(a x + b|x^*,a,b)] = E [a^2 \text{Var}(x^*)] \\
            &= E [a^2 \sigma_{x^*}^2] \\
            &= \sigma_{x^*}^2 E [a^2]; a^2 \sim \chi^2 (\text{d.o.f}=1)\\
            &= \sigma_{x^*}^2 (\mu_a^2 + \sigma_a^2)
        \end{aligned}
    \end{equation}

    Equation \ref{eq:data_unc} captures the variance that is influenced by the error in cardinality $x^*$ and the sensitivity of the cost model to such errors. This component has a direct relationship with the uncertainty in the input data ($\sigma_{x^*}^2$) on the one hand and the mean and variance of the slope (i.e. the first derivative) ($\mu_a^2 + \sigma_a^2$) of $f_\theta(x)$ on the other. The term $\mu_a^2 + \sigma_a^2$ determines the level of sensitivity to cardinality error. \textit{Therefore, mapping the term $E[\text{Var} (f_\theta(x)|x^*,\theta)]$ to the robustness of the plan is justified as it captures both the error rooted in the input cardinality and the sensitivity of the plan to this error.} Note that this component is reduced to zero if $\sigma_{x^*}^2 = 0$ or $\mu_a^2 + \sigma_a^2 = 0$.
    
    \begin{equation}
        \begin{aligned}
            \text{Var} (E [f_\theta(x) |x^*,\theta]) &= \text{Var} (E[a.x + b|x^*,a,b]) = \text{Var} (a.E[x^*] + b) \\
            &= \text{Var} (a.\mu_{x^*} + b) \\
            &= \mu_{x^*}^2 . \text{Var}(a) + \text{Var}(b) + 2 \text{Cov} (a.\mu_{x^*},b) \\
            &= \mu_{x^*}^2 .\sigma^2_a + \sigma^2_b + 2 \mu_{x^*}. \text{Cov} (a, b)
            \label{eq:model_unc}
        \end{aligned}
    \end{equation}
    
    Equation \ref{eq:model_unc} captures the uncertainty rooted in the model parameters $\sigma_a^2$ and $\sigma_b^2$, and their covariance $\text{Cov} (a, b)$. This is the uncertainty rooted in the limitations of the cost model. \textit{Therefore, mapping the term $\text{Var} (E [f_\theta(x) |x^*,\theta])$ to the uncertainty rooted in model parameters is justified.} Note that the term $\text{Var} (E [f_\theta(x) |x^*,\theta])$ is reduced to zero if and only if $\sigma_a^2 = 0$, $\sigma_b^2 = 0$, and $\text{Cov}(a, b) = 0$. In such a scenario, the total variance of the plan's cost estimate will be determined by the cardinality errors and the sensitivity of the cost to those errors.  

    Note that the assumption of a normal distribution for $a$, $b$, and $x^*$ is made to facilitate the discussion and that it is inconsequential to the conclusions. The same applies to the assumption of linearity.
    \end{proof}

    \subsection{Extended Discussion on the Formulation of SOR}
    \label{sec:appendixA.2}
    
    To instantiate SOR with a Gaussian prior, let us rewrite Equation \ref{eq:SOR_dist} as:
        \[
        \delta = C_x - C_y \sim \mathcal{N}(\mu_\delta,\sigma_\delta^2)
        \]
    where $\mu_\delta = \mu_x - \mu_y, \sigma_\delta^2 = \sigma_x^2 + \sigma_y^2$
    The probability density function $\mathcal{F}(\delta)$ for a variable $\delta$ with normal distribution $\mathcal{N}(\mu_\delta,\sigma_\delta^2)$ is:
    \[
    \mathcal{F}(\delta) = \frac{1}{\sqrt{2\pi\sigma_\delta^2}} e^{-\frac{(\delta - \mu_\delta)^2}{2\sigma^2}}
    \]
    
    By substituting $\delta$, $\mu_\delta$ and $\sigma_\delta^2$ this equation can be expanded as:
    \[
    \mathcal{F}(\delta) = \frac{1}{\sqrt{2\pi(\sigma_x^2+\sigma_y^2)}} e^{-\frac{(C_x - C_y - (\mu_x-\mu_y))^2}{2(\sigma_x^2+\sigma_y^2)}}
    \]

    The risk of suboptimality of plan $x$ over plan $y$ (i.e. $R(x,y)$) can be computed as:
    \[
    R(x,y) = P(C_x > C_y) = P(\delta > 0) = \int_{0}^{\infty} \mathcal{F}(\delta) d\delta
    \]

    where as demonstrated above, $\mathcal{F}(\delta)$ is a function of $C_x$, $C_y$, $\mu_x$, $\mu_y$, $\sigma_x$, and $\sigma_y$. 

    This demonstrates that $\text{SOR}(p_i)$, which is the average risk of suboptimality of plan $p_i$ as given in Equation \ref{eq:SOR}, is a function of the expected value and variance of cost (and hence robustness) of all plans in the search space.

    \subsection{Hint Sets Used for Plan Generation}
    \label{hint-sets}
      Inspired by the finding reported by Bao \cite{marcus_bao_2021}, these hint sets were used which include the most influential among the total 46 hint sets: 
        [disable nljn], [disable nljn, disable iscan], [disable hsjn], [disable hsjn, disable iscan], [disable mgjn], [disable mgjn, disable iscan], [disable nljn, disable mgjn ], [disable nljn, disable mgjn, disable iscan], [disable nljn, disable hsjn ], [disable nljn, disable hsjn, disable iscan], [disable mgjn, disable hsjn ], [disable mgjn, disable hsjn, disable iscan]
\end{appendices}

\end{document}